\documentclass[preprint]{elsarticle}

\usepackage[english]{babel}

\usepackage{ifpdf}

\usepackage{lineno,hyperref} 

\usepackage{graphicx}
    \graphicspath{{./figures/}}
\usepackage{color}
\usepackage{pgf, tikz, pgfplots}
\usetikzlibrary{shapes, arrows, automata}
\usepackage{placeins}

\usepackage[cmex10]{amsmath}
\usepackage{amsfonts, amssymb, amsthm}
\usepackage{mathrsfs}

\usepackage[]{algorithm2e}

\usepackage{enumerate}
\usepackage{multirow}
\usepackage{rotating}
\usepackage{subcaption}
\usepackage{needspace}

\input{mySymbol.sty}

\def\digitsize{0.99} 
\def\threeplotsize{0.99} 

\definecolor{pennpurple}{rgb}{0.537,0,0.510} 

\newtheorem{proposition}{\hspace{0pt}\bf Proposition}
\newtheorem{remark}{\hspace{0pt}\bf Remark}

\renewcommand \blue[1]        {{\color{black}#1}}

\journal{Signal Processing}

\begin{document}

\begin{frontmatter}

\title{Rethinking Sketching as Sampling:\\A Graph Signal Processing Approach\tnoteref{funding,conferences}}
\tnotetext[funding]{Work in this paper is supported by NSF CCF 1750428, NSF ECCS 1809356, NSF CCF 1717120, ARO W911NF1710438 and Spanish MINECO grants No TEC2013-41604-R and TEC2016-75361-R.}
\tnotetext[conferences]{Part of the results in this paper were presented at the \textit{2016 Asilomar Conference of Signals, Systems and Computers}~\cite{FGAMGMAR_asilomar16} and the \textit{2016 IEEE GobalSIP Conference}~\cite{FGAMGMAR_globalsip16}.}

\author[upenn]{Fernando Gama\corref{cor1}}
\ead{fgama@seas.upenn.edu}
\cortext[cor1]{Corresponding author}

\author[urjc]{Antonio G. Marques\fnref{myfootnote}}
\ead{antonio.garcia.marques@urjc.es}

\author[rochester]{Gonzalo Mateos}
\ead{gmateosb@ece.rochester.edu}

\author[upenn]{Alejandro Ribeiro}
\ead{aribeiro@seas.upenn.edu}

\address[upenn]{Dept. of Electrical and Systems Eng., Univ. of Pennsylvania, Philadelphia, USA}
\address[urjc]{Dept. of Signal Theory and Comms., King Juan Carlos Univ., Madrid, Spain}
\address[rochester]{Dept. of Electrical and Computer Eng., Univ. of Rochester, Rochester, USA}

\begin{abstract}
Sampling of signals belonging to a low-dimensional subspace has well-documented merits for dimensionality reduction, limited memory storage, and online processing of streaming network data. \blue{When the subspace is known, these signals can be modeled as bandlimited graph signals.} Most existing sampling methods are designed to minimize the error incurred when reconstructing the original signal from its samples. Oftentimes these parsimonious signals serve as inputs to computationally-intensive linear operators. Hence, interest shifts from reconstructing the signal itself towards approximating the output of the prescribed linear operator efficiently. In this context, we propose a novel sampling scheme that leverages \blue{graph signal processing, exploiting} the low-dimensional \blue{(bandlimited)} structure of the input as well as the transformation whose output we wish to approximate. We formulate problems to jointly optimize sample selection and a sketch of the target linear transformation, so when the latter is applied to the sampled input signal the result is close to the desired output. Similar sketching as sampling ideas are also shown effective in the context of linear inverse problems. Because these designs are carried out off line, the resulting sampling plus reduced-complexity processing pipeline is particularly useful for data that are acquired or processed in a sequential fashion, where the linear operator has to be applied fast and repeatedly to successive inputs or response signals. Numerical tests showing the effectiveness of the proposed algorithms include classification of handwritten digits from as few as 20 out of 784 pixels in the input images and selection of sensors from a network deployed to carry out a distributed parameter estimation task.
\end{abstract}


\end{frontmatter}


%
\section{Introduction} \label{sec_intro}


The complexity of modern datasets calls for new tools capable of analyzing and processing signals supported on irregular domains. A key principle to achieve this goal is to explicitly account for the intrinsic structure of the domain where the data resides. This is oftentimes achieved through a parsimonious description of the data, for instance modeling them as belonging to a known (or otherwise learnt) lower-dimensional subspace. Moreover, these data can be further processed through a linear transform to estimate a quantity of interest \cite{Moulines95-SubspaceFIR, Dai09-SubspaceReconstruction}, or to obtain an alternative, more useful representation \cite{Akcay14-SubspaceFrequency, Giampouras17-SubspaceIncomplete} among other tasks \cite{Yan07-SubspaceGraph, Elhamifar13-SubspaceClustering}. Linear models are ubiquitous in science and engineering, due in part to their generality, conceptual simplicity, and mathematical tractability. Along with heterogeneity and lack of regularity, data are increasingly high dimensional and this curse of dimensionality not only raises statistical challenges, but also major computational hurdles even for linear models~\cite{slavakis14-mag}. In particular, these limiting factors can hinder processing of streaming data, where say a massive linear operator has to be repeatedly  and efficiently applied to a sequence of input signals \cite{berberidis16}. These Big Data challenges motivated a recent body of work collectively addressing so-termed \emph{sketching} problems~\cite{boutsidis09-colsel,mahoney11,woodruff14}, which seek computationally-efficient solutions to a subset of (typically inverse) linear problems. The basic idea is to \emph{draw a sketch} of the linear model such that the resulting linear transform is lower dimensional, while still offering quantifiable approximation error guarantees. To this end, a fat random projection matrix is designed to pre-multiply and reduce the dimensionality of the linear operator matrix, in such way that the resulting matrix sketch still captures the quintessential structure of the model. The input vector has to be adapted to the sketched operator as well, and to that end the same random projections are applied to the signal in a way often agnostic to the input statistics.

Although random projection methods offer an elegant dimensionality reduction alternative for several Big Data problems,  they face some shortcomings: i) sketching each new input signal entails a nontrivial computational cost, which can be a bottleneck in streaming applications; ii) the design of the random projection matrix does not take into account any a priori information on the input \blue{(known subspace)}; and iii) the guarantees offered are probabilistic. 
Alternatively one can think of reducing complexity by simply retaining a few samples of each input. \blue{A signal belonging to a known subspace can be modeled as a \emph{bandlimited} graph signal \cite{shuman13,sandryhaila14}. In the context of graph signal processing (GSP), sampling of bandlimited graph signals} has been thoroughly studied \cite{shuman13,sandryhaila14}, giving rise to several noteworthy sampling schemes \cite{chen15,marques16,tsitsvero16,anis16}. Leveraging these advances along with the concept of stationarity \cite{girault15, perraudin16, marques2016stationaryTSP16} offers novel insights to design sampling patterns accounting for the signal statistics, that can be applied at negligible online computational cost to a stream of inputs. However, most existing sampling methods are designed with the objective of reconstructing the original graph signal, and do not account for subsequent processing the signal may undergo; see \cite{gray06,Geert,chepuri16} for a few recent exceptions.

In this sketching context and towards reducing the online computational cost of obtaining the solution to a linear problem, we leverage GSP results and propose a novel sampling scheme for signals that belong to a \emph{known} low-dimensional subspace. Different from most existing sampling approaches, our design explicitly accounts for the transformation whose output we wish to approximate. By exploiting the stationary nature of the sequence of inputs, we shift the computational burden to the off-line phase where both the sampling pattern and the sketch of the linear transformation are designed. After doing this only once, the online phase merely consists of repeatedly selecting the signal values dictated by the sampling pattern and processing this stream of samples using the sketch of the linear transformation.

In Section \ref{sec_preliminaries} we introduce the mathematical formulation of the direct and inverse linear sketching problems as well as the assumptions on the input signals. Then, we proceed to present the solutions for the direct and inverse problems in Section \ref{sec_forward}. In both cases, we obtain first a closed-form expression for the optimal reduced linear transform as a function of the selected samples of the signal. Then we use that expression to obtain an equivalent optimization problem on the selection of samples, that turns out to be a Semidefinite Program (SDP) modulo binary constraints that arise naturally from the sample (node) selection problem. Section \ref{sec_heuristics} discusses a number of heuristics to obtain tractable solutions to the binary optimization. In Section \ref{sec_sims} we apply this framework to the problem of estimating the graph frequency components of a graph signal in a fast and efficient fashion, as well as to the problems of selecting sensors for parameter estimation, classifying handwritten digits and attributing texts to their corresponding author. Finally, conclusions are drawn in Section \ref{sec_conclusions}.

\paragraph{Notation} Generically, the entries of a matrix $\mathbf{X}$ and a (column) vector $\mathbf{x}$ will be denoted as $X_{ij}$ and $x_i$. The notation $^T$ and $^H$ stands for transpose and transpose conjugate, respectively, and the superscript $^{\dagger}$ denotes pseudoinverse; $\mathbf{0}$ is the all-zero vector and $\bbone$ is the all-one vector; and the $\ell_0$ pseudo norm $\| \bbX\|_{0}$ equals the number of nonzero entries in $\bbX$. For a vector $\bbx$, $\diag(\mathbf{x})$ is a diagonal matrix with the $(i,i)$th entry equal to $x_{i}$; when applied to a matrix, $\diag(\bbX)$ is a vector with the diagonal elements of $\bbX$. For vectors $\bbx, \bby \in \reals^{n}$ we adopt the partial ordering $\preceq$ defined with respect to the positive orthant $\reals^{n}_{+}$ by which a $\bbx \preceq \bby$ if and only if $x_{i} \leq y_{i}$ for all $i=1,\ldots,n$. For symmetric matrices $\bbX,\bbY \in \reals^{n \times n}$, the partial ordering $\preceq$ is adopted with respect to the semidefinite cone, by which $\bbX \preceq \bbY$ if and only if $\bbY - \bbX$ is positive semi-definite.

%
\section{Sketching of bandlimited signals} \label{sec_preliminaries}


Our results draw inspiration from the existing literature for sampling of bandlimited graph signals. So we start our discussion in Section~\ref{sssec_gsp} by defining sampling in the GSP context, and explaining how these ideas extend to more general signals that belong to lower-dimensional subspaces. Then in Sections \ref{sssec_prob_direct}  and \ref{sssec_prob_inverse}  we present, respectively, the direct and inverse formulation of our sketching as sampling problems. GSP applications and motivating examples that involve linear processing of network data are presented in Section \ref{sec_sims}.

\subsection{Graph signal processing}\label{sssec_gsp}

Let $\ccalG=(\ccalV,\ccalE,\ccalW)$ be a graph described by a set of $n$ nodes $\ccalV$, a set $\ccalE$ of edges $(i,j)$ and a weight function $\ccalW:\ccalE \to \reals$ that assigns weights to the directed edges. Associated with the graph we have a shift operator $\bbS \in \reals^{n \times n}$ which we define as a matrix sharing the sparsity pattern of the graph so that $[\bbS]_{i,j}=0$ for all $i\neq j$ such that $(j,i) \notin \ccalE$ \cite{puschel08}. The shift operator is assumed normal so that there exists a matrix of eigenvectors $\bbV = [\bbv_{1},\ldots,\bbv_n]$ and a diagonal matrix $\bbLam=\diag(\lam_1,\ldots,\lam_n)$ such that

\begin{equation}\label{eqn_shift_operator_decomposition}
   \bbS \  = \ \bbV \bbLam \bbV^H .
\end{equation}
We consider realizations $\bbx=[x_{1}, \ldots, x_{n}]^T  \in\reals^n$ of a random signal with zero mean $\mbE[\bbx]=\bbzero $ and covariance matrix $\bbR_{x} := \mbE[\bbx\bbx^T]$. The random signal $\bbx$ is interpreted as being supported on $\ccalG$ in the sense that components $x_i$ of $\bbx$ are associated with node $i$ of $\ccalG$. The graph is intended as a descriptor of the relationship between components of the signal $\bbx$. The signal $\bbx$ is said to be stationary on the graph if the eigenvectors of the shift operator $\bbS$ and the eigenvectors of the covariance matrix $\bbR_x$ are the same \cite{girault15, marques2016stationaryTSP16, perraudin16}. It follows that there exists a diagonal matrix $\tbR_x\succeq \mathbf{0}$ that allows us to write

\begin{equation}\label{eqn_covariance}
   \bbR_{x} \ := \ \mbE[\bbx\bbx^T] 
            \  = \ \bbV \tbR_x \bbV^H .
\end{equation}
The diagonal entry $[\tbR_x]_{ii} = \tdr_i$  is the eigenvalue associated with eigenvector $\bbv_i$. Without loss of generality we assume eigenvalues are ordered so that $\tdr_i \geq \tdr_j$ for $i\leq j$. Crucially, we assume that exactly $k \leq n$ eigenvalues are nonzero. This implies that if we define the subspace projection $\tbx:=\bbV^H \bbx$, its last $n-k$ elements are almost surely zero. Therefore, upon defining the vector $\tbx_{k}:=[\tdx_1, \ldots, \tdx_k]^T \in \reals^k$ containing the first $k$ elements of $\tbx$ it holds with probability $1$ that 

\begin{equation}\label{eqn_projection}
   \tbx \ := \ \bbV^H \bbx
        \  = \ [\tbx_{k}; \bbzero_{n-k}]^T .
\end{equation}
Stationary graph signals can arise, e.g., when considering diffusion processes on the graph and they will often have spectral profiles with a few dominant eigenvalues \cite{marques2016stationaryTSP16, Gama19-GLLN, Gama19-Control}. Stationary graph signals are important in this paper because they allow for interesting interpretations and natural connections to the sampling of bandlimited graph signals \cite{chen15, marques16, tsitsvero16, anis16, gray06, Geert, chepuri16} -- see Section \ref{sec_forward}. That said, techniques and results apply for as long as \eqref{eqn_projection} holds whether there is a graph that supports the signal or not.

Observe for future reference that since $\bbV\tbx = \bbV\bbV^H\bbx = \bbx$ we can undo the projection with multiplication by the eigenvector matrix $\bbV$. This can be simplified because, as per  \eqref{eqn_projection}, only the first $k$ entries of $\tbx$ are of interest. Define then the (tall) matrix $\bbV_k = [\bbv_{1},\ldots,\bbv_k] \in \mbC^{n \times k}$ containing the first $k$ eigenvectors of $\bbR_x$. With this definition we can write $\tbx_k = \bbV_k^H \bbx$ and 

\begin{equation}\label{eqn_recovery}
   \bbx \ = \ \bbV_k \tbx_k 
        \ = \ \bbV_k (\bbV_k^H \bbx).
\end{equation}
When a process is such that realizations lie in a $k$-dimensional subspace as specified in \eqref{eqn_projection} we say that it is $k$-bandlimited  or simply bandlimited if $k$ is understood. We emphasize that in this definition $\bbV$ is an arbitrary orthonormal matrix which need not be associated with a shift operator as in \eqref{eqn_shift_operator_decomposition} -- notwithstanding the fact that it will be associated with a graph in some applications. Our goal here is to use this property to sketch the computation of linear transformations of $\bbx$ or to sketch the solution of linear systems of equations involving $\bbx$ as we explain in Sections \ref{sssec_prob_direct} and \ref{sssec_prob_inverse}. 

%
\begin{figure}\centering
\begin{minipage}[c]{0.48\linewidth}
    \def \thisplotscale {0.45}
\def \unit {\thisplotscale cm}

\tikzstyle {block} = [draw,
                      rectangle, 
                      fill = blue!10, 
                      minimum height = 2*\unit]

\tikzstyle {long block}  = [block,
                            minimum width = 4*\unit]

\tikzstyle {short block} = [block,
                            minimum width = 2*\unit]

\tikzstyle {add block}   = [draw,
                            circle,
                            minimum width = 0.9*\unit]

{\footnotesize\begin{tikzpicture}[x = 1*\unit, y=1*\unit, font=\footnotesize]

%
\path (0,0) node [long block] (H) {$\bbH$};
\path (H.east) ++ (0,-3) node [short block] (H2) {$\bbH\bbC^T$};
\path (H.west |- H2) node [short block] (C) {$\bbC$};
\path (H.west) ++ (-3,0) node [add block] (addition) {$+$};

%
\path [draw, -stealth] (addition.east) -- (H.west);
\path [draw, -stealth] (H.east) -- (5,0)  node  [above] (y) {$\bby$};
\path [draw, -stealth] (addition.south) -- (addition.south |- C.west) -- (C.west);
\path [draw, -stealth] (C.east) -- (H2.west);
\path [draw, -stealth] (H2.east) -- (H2.east -| y) node [above] {$\hby$};
\path [draw, -stealth] (addition) ++ (0,2) node [right] {$\bbw$} -- (addition.north);

\path [draw, -stealth] (addition) ++ (-2,0) node [above] {\bbx} -- (addition.west);

\end{tikzpicture}}    
\end{minipage}\quad
\begin{minipage}[c]{0.48\linewidth}
    \def \thisplotscale {0.45}
\def \unit {\thisplotscale cm}

\tikzstyle {block} = [draw,
                      rectangle, 
                      fill = red!10, 
                      minimum height = 2*\unit]

\tikzstyle {long block}  = [block,
                            minimum width = 4*\unit]

\tikzstyle {short block} = [block,
                            minimum width = 2*\unit]

\tikzstyle {add block}   = [draw,
                            circle,
                            minimum width = 0.9*\unit]

{\footnotesize\begin{tikzpicture}[x = 1*\unit, y=1*\unit, font=\footnotesize]

%
\path (0,0) node [long block] (H) {$\bbH$};
\path (H.east) ++ (0,-3) node [short block] (H2) {$\bbH_s$};
\path (H.west |- H2) node [short block] (C) {$\bbC$};
\path (H.west) ++ (-3,0) node [add block] (addition) {$+$};

%
\path [draw, -stealth] (addition.east) -- (H.west);
\path [draw, -stealth] (H.east) -- (5,0)  node  [above] (y) {$\bby$};
\path [draw, -stealth] (addition.south) -- (addition.south |- C.west) -- (C.west);
\path [draw, -stealth] (C.east) -- (H2.west);
\path [draw, -stealth] (H2.east) -- (H2.east -| y) node [above] {$\hby$};
\path [draw, -stealth] (addition) ++ (0,2) node [right] {$\bbw$} -- (addition.north);

\path [draw, -stealth] (addition) ++ (-2,0) node [above] {\bbx} -- (addition.west);

\end{tikzpicture}}
\end{minipage}
    \caption{Direct sketching problem. We observe realizations $\bbx+\bbw$ and want to estimate the output $\bby = \bbH \bbx$ with a reduced complexity pipeline. (Left) We sample the incoming observation $\bbx_{s} = \bbC (\bbx + \bbw)$, and multiply it by the corresponding columns $\bbH_{s} = \bbH \bbC^{T}$ [cf. \eqref{eqn_operator_direct_sketching}]; the appropriate samples for all incoming observations are determined by \eqref{eqn_operator_direct_sketching_optimal_matrix}. (Right) Since the multiplication by $\bbH_{s}$ is the same for all incoming $\bbx_{s}$ [cf. \eqref{eqn_direct_sketching}] we can design, off-line, an arbitrary matrix $\bbH_{s}$ that improves the performance, as per \eqref{eqn_direct_sketching_optimal_matrix}.}
    \label{fig_forward} 
\end{figure}

%
\subsection{Direct sketching} \label{sssec_prob_direct} 

The direct sketching problem is illustrated in Figure \ref{fig_forward}. Consider a noise vector $\bbw\in\reals^n$ with zero mean $\mbE[\bbw]=\bbzero $ and covariance matrix $\bbR_{w} := \mbE[\bbw\bbw^T]$. We observe realizations $\bbx+\bbw$ and want to estimate the matrix-vector product $\bby = \bbH \bbx$ for a matrix $\bbH\in \reals^{m\times n}$. Estimating this product requires $O(mn)$ operations. The motivation for sketching algorithms is to devise alternative computation methods requiring a (much) smaller number of operations in settings where multiplication by the matrix $\bbH$ is to be carried out for a large number of realizations $\bbx$. In this paper we leverage the bandlimitedness of $\bbx$ to design sampling algorithms to achieve this goal.

Formally, we define binary selection matrices $\bbC$ of dimension $p\times n$ as those that belong to the set 

\begin{align}\label{eqn_sampling_matrix}
   \ccalC := \{ \bbC\in\{0,1\}^{p\times n} :\  
                    \bbC\bbone = \bbone, \
                    \bbC^T\bbone\preceq\bbone\ 
                 \}.
\end{align}
The restrictions in \eqref{eqn_sampling_matrix} are such that each row of $\bbC$ contains exactly one nonzero element and that no column of $\bbC$ contains more than one nonzero entry. Thus, the product vector $\bbx_s:=\bbC\bbx \in\reals^p$ samples (selects) $p$ elements of $\bbx$ that we use to estimate the product $\bby = \bbH \bbx$. In doing so we not only need to select entries of $\bbx$ but columns of $\bbH$. This is achieved by computing the product $\bbH_s:=\bbH\bbC^T \in \reals^{m\times p}$ and we therefore choose to estimate the product $\bby = \bbH \bbx$ with the product 

\begin{align}\label{eqn_operator_direct_sketching}
   \hby \ := \ \bbH_s \bbx_s
        \ := \ \bbH\bbC^T \bbC(\bbx+\bbw) .
\end{align}
Implementing \eqref{eqn_operator_direct_sketching} requires $O(mp)$ operations. Adopting the mean squared error (MSE) as a figure of merit, the optimal sampling matrix $\bbC \in \ccalC$ is the solution to the minimum (M)MSE problem comparing $\hby$ in \eqref{eqn_operator_direct_sketching} to the desired response $\bby = \bbH\bbx$, namely 

\begin{align}\label{eqn_operator_direct_sketching_optimal_matrix}
   \bbC^* := \argmin_{\bbC \in \ccalC} 
            \E{ \big\| \bbH\bbC^T \bbC(\bbx+\bbw) - \bbH \bbx) \big\|_2^2}.
\end{align}
Observe that selection matrices $\bbC\in\ccalC$ have rank $p$ and satisfy $\bbC\bbC^T=\bbI$, the $p$-dimensional identity matrix. It is also readily verified that $\bbC^T\bbC=\diag(\bbc)$ with the vector $\bbc\in\{0,1\}^n$ having entries $c_i = 1$ if and only if the $i$th column of $\bbC$ contains a nonzero entry. Thus, the product $\bbC^T\bbC$ is a diagonal matrix in which the $i$th entry is nonzero if and only if the $i$th entry of $\bbx$ is selected in the sampled vector $\bbx_s:=\bbC\bbx$. In particular, there is a bijective correspondence between the matrix $\bbC$ and the vector $\bbc$ modulo an arbitrary ordering of the rows of $\bbC$. In turn, this implies that choosing $\bbC \in \ccalC$ in \eqref{eqn_operator_direct_sketching_optimal_matrix} is equivalent to choosing $\bbc\in\{0,1\}^n$ with $\bbc^T\bbone=p$. We take advantage of this fact in the algorithmic developments in Sections \ref{sec_forward} and \ref{sec_heuristics}.

In \eqref{eqn_operator_direct_sketching}, the sampled signal $\bbx_s$ is multiplied by the matrix $\bbH_s:=\bbH\bbC^T$. The restriction to have the matrix $\bbH_s$ be a sampled version of $\bbH$ is unnecessary in cases where it is possible to compute an arbitrary matrix off line. This motivates an alternative formulation where we estimate $\bby$ as

\begin{align}\label{eqn_direct_sketching}
   \hby \ := \ \bbH_s \bbx_s
        \ := \ \bbH_s \bbC(\bbx+\bbw) .
\end{align}
The matrices $\bbH_s$ and $\bbC$ can now be jointly designed as the solution to the MMSE optimization problem 

\begin{equation}\label{eqn_direct_sketching_optimal_matrix}
   \{\bbC^*,\bbH_s^*\}  := 
       \argmin_{\bbC\in\ccalC,\bbH_s} 
            \E{ \big\| \bbH_s \bbC (\bbx+\bbw) - \bbH \bbx  \big\|_2^2}.
\end{equation}
To differentiate \eqref{eqn_operator_direct_sketching_optimal_matrix} from \eqref{eqn_direct_sketching_optimal_matrix} we refer to the former as {\it operator} sketching since it relies on sampling the operator $\bbH$ that operates on the vector $\bbx$. In \eqref{eqn_direct_sketching} we refer to $\bbH_s$ as the sketching matrix which, as per \eqref{eqn_direct_sketching_optimal_matrix}, is jointly chosen with the sampling matrix $\bbC$. In either case we expect that sampling $p \approx k$ entries of $\bbx$ should lead to good approximations of $\bby$ given the assumption that $\bbx$ is $k$-bandlimited. We will see that $p=k$ suffices in the absence of noise (Proposition \ref{prop_noiseless_forward}) and that making $p>k$ helps to reduce the effects of noise otherwise (Proposition \ref{prop_noisy_forward}). We point out  that solving \eqref{eqn_operator_direct_sketching_optimal_matrix} or \eqref{eqn_direct_sketching_optimal_matrix} is intractable because of their nonconvex objectives and the binary nature of the matrices $\bbC\in\ccalC$. Heuristics for approximate solution with manageable computational cost are presented in Section \ref{sec_heuristics}. While tractable, these heuristics still entail a significant computation cost. This is justified when solving a large number of estimation tasks. See Section \ref{sec_sims} for concrete examples.

%
\begin{figure}\centering
\begin{minipage}[c]{0.48\linewidth}
    \def \thisplotscale {0.45}
\def \unit {\thisplotscale cm}

\tikzstyle {block} = [draw,
                      rectangle, 
                      fill = blue!10, 
                      minimum height = 2*\unit]

\tikzstyle {long block}  = [block,
                            minimum width = 4*\unit]

\tikzstyle {short block} = [block,
                            minimum width = 2*\unit]

\tikzstyle {add block}   = [draw,
                            circle,
                            minimum width = 0.9*\unit]

{\footnotesize\begin{tikzpicture}[x = 1*\unit, y=1*\unit, font=\footnotesize]

%
\path (0,0) node [long block] (H) {$\bbH^{T}$};
\path (H.east) ++ (0,-3) node [short block] (C) {$\bbC$};
\path (H.west |- C) node [short block] (H2) {$\bbH_{s}(\bbH, \bbC) $};
\path (H.east) ++ (3,0) node [add block] (addition) {$+$};

%
\path [draw, -stealth] (H.west) ++ (-3,0) node [above] (x) {$\bby$} -- (H.west);
\path [draw, -stealth] (H.east) -- (addition.west);
\path (H.east) ++ (1,0) node  [above] {$\bbx$};
\path [draw, -stealth] (addition.south) -- (addition.south |- C) -- (C.east);
\path [draw, -stealth] (C.west) -- (H2.east);
\path [draw, -stealth] (H2.west) -- (H2.west -| x) node [above] {$\hby$};
\path [draw, -stealth] (addition) ++ (2,0) node [above] {$\bbw$} -- (addition.east);

\end{tikzpicture}}    
\end{minipage}\quad
\begin{minipage}[c]{0.48\linewidth}
    \def \thisplotscale {0.45}
\def \unit {\thisplotscale cm}

\tikzstyle {block} = [draw,
                      rectangle, 
                      fill = red!10, 
                      minimum height = 2*\unit]

\tikzstyle {long block}  = [block,
                            minimum width = 4*\unit]

\tikzstyle {short block} = [block,
                            minimum width = 2*\unit]

\tikzstyle {add block}   = [draw,
                            circle,
                            minimum width = 0.9*\unit]

{\footnotesize\begin{tikzpicture}[x = 1*\unit, y=1*\unit, font=\footnotesize]

%
\path (0,0) node [long block] (H) {$\bbH^{T}$};
\path (H.east) ++ (0,-3) node [short block] (C) {$\bbC$};
\path (H.west |- C) node [short block] (H2) {$\bbH_s$};
\path (H.east) ++ (3,0) node [add block] (addition) {$+$};

%
\path [draw, -stealth] (H.west) ++ (-3,0) node [above] (x) {$\bby$} -- (H.west);
\path [draw, -stealth] (H.east) -- (addition.west);
\path (H.east) ++ (1,0) node  [above] {$\bbx$};
\path [draw, -stealth] (addition.south) -- (addition.south |- C) -- (C.east);
\path [draw, -stealth] (C.west) -- (H2.east);
\path [draw, -stealth] (H2.west) -- (H2.west -| x) node [above] {$\hby$};
\path [draw, -stealth] (addition) ++ (2,0) node [above] {$\bbw$} -- (addition.east);

\end{tikzpicture}}
\end{minipage}
	\caption{Inverse sketching problem. Reduce the computational cost of solving a least squares estimation problem $\bbx = \bbH^{T} \bby$. (Left) Given observations of $\bbx+\bbw$, we sample them $\bbx_{s} = \bbC(\bbx + \bbw)$ and solve the linear regression problem of the reduced system $ \bbC \bbx = \bbC \bbH^{T} \bby$ by computing the least squares solution $\bbH_{s}(\bbH, \bbC) = (\bbH\bbC^T\bbC\bbH^{T})^{-1} \bbH\bbC^T$ and multiplying it by the sampled observations [cf. \eqref{eqn_operator_inverse_sketching}]; the optimal samples $\bbC$ are designed by solving \eqref{eqn_operator_inverse_sketching_optimal_matrix}. (Right) Likewise, instead of solving the least squares problem on the reduced matrix, we can design an entirely new smaller matrix $\bbH_{s}$ that acts on the sampled observations [cf. \eqref{eqn_inverse_sketching}], jointly designing the sampling pattern $\bbC$ and the sketch $\bbH_{s}$ as per \eqref{eqn_inverse_sketching_optimal_matrix}.}
	\label{fig_backward}
\end{figure}

\subsection{Inverse sketching} \label{sssec_prob_inverse} 

The inverse sketching problem seeks to solve a least squares estimation problem with reduced computational cost. As in the case of the direct sketching problem we exploit the bandlimitedness of $\bbx$ to propose the sampling schemes illustrated in Figure \ref{fig_backward}. Formally, we consider the system $\bbx=\bbH^T\bby$ and want to estimate $\bby$ from observations of the form $\bbx+\bbw$. As in Section \ref{sssec_prob_direct}, the signal $\bbx\in\reals^n$ is $k$-bandlimited as described in \eqref{eqn_covariance}-\eqref{eqn_recovery} and the noise $\bbw\in\reals^n$ is zero mean with covariance $\bbR_{w} := \mbE[\bbw\bbw^T]$. The signal of interest is $\bby\in\reals^m$ and the matrix that relates $\bbx$ to $\bby$ is $\bbH\in\reals^{m\times n}$. The solution to this least squares problem is to make $\hby = (\bbH \bbH^{T})^{-1}\bbH (\bbx+\bbw)$ which requires $O(mn)$ operations if the matrix $\bbA_\mathrm{LS}=(\bbH \bbH^{T})^{-1}\bbH$ is computed off line or $O(m^{2} n)$ operations if the matrix $\bbH \bbH^{T}$ is computed online. 

To reduce the cost of computing the least squares estimate we consider sampling matrices $\bbC\in\ccalC$ as defined in \eqref{eqn_sampling_matrix}. Thus, the sampled vector $\bbx_s := \bbC(\bbx+\bbw)$ is one that selects $p$ entries of the observation $\bbx+\bbw$. Sampling results in a reduced observation model and leads to the least square estimate

\begin{align}\label{eqn_operator_inverse_sketching}
   \hby \ := \ \bbH_s \bbx_s 
        \ := \ (\bbH\bbC^T\bbC\bbH^{T})^{-1} \bbH\bbC^T \bbC(\bbx+\bbw)
\end{align}
where we have defined the estimation matrix $\bbH_s:=(\bbH\bbC^T\bbC\bbH^{T})^{-1} \bbH\bbC^T$. The computational cost of implementing \eqref{eqn_operator_inverse_sketching} is $O(mp)$ operations if the matrix $\bbH_s$ is computed off line or $O(m^{2}p)$ operations if the matrix $\bbH_s$ is computed online. We seek the optimal sampling matrix that minimizes the MSE

\begin{align}\label{eqn_operator_inverse_sketching_optimal_matrix}
   \bbC^* := 
       \argmin_{\bbC\in\ccalC} 
            \E{ \big\| \bbH^{T} 
                       (\bbH\bbC^T\bbC\bbH^{T})^{-1} \bbH\bbC^T 
                       \bbC (\bbx+\bbw) 
                       -\bbx \big\|_2^2}.
\end{align}
As in \eqref{eqn_operator_direct_sketching_optimal_matrix}, restricting $\bbH_s$ in \eqref{eqn_operator_inverse_sketching} to be a sampled version of $\bbH$ is unnecessary if $\bbH_s$ is to be computed off line. In such case we focus on estimates of the form %

\begin{align}\label{eqn_inverse_sketching}
   \hby \ := \ \bbH_s \bbx_s
        \ := \ \bbH_s \bbC(\bbx+\bbw)
\end{align}
where the matrix $\bbH_s\in\reals^{m\times p}$ is an arbitrary matrix that we select jointly with $\bbC$ to minimize the MSE,

\begin{equation}\label{eqn_inverse_sketching_optimal_matrix}
   \{\bbC^*,\bbH_s^*\} := 
       \argmin_{\bbC\in\ccalC,\bbH_s} 
            \E{ \big\| \bbH^{T} \bbH_s \bbC (\bbx+\bbw) -\bbx \big\|_2^2}.
\end{equation}
We refer to \eqref{eqn_operator_inverse_sketching}-\eqref{eqn_operator_inverse_sketching_optimal_matrix} as the inverse {\it operator} sketching problem and to \eqref{eqn_inverse_sketching}-\eqref{eqn_inverse_sketching_optimal_matrix} as the inverse sketching problem. They differ in that the matrix $\bbH_s$ is arbitrary and jointly optimized with $\bbC$ in \eqref{eqn_inverse_sketching_optimal_matrix} whereas it is restricted to be of the form $\bbH_s:=(\bbH\bbC^T\bbC\bbH^{T})^{-1} \bbH\bbC^T$ in \eqref{eqn_operator_inverse_sketching_optimal_matrix}. Inverse sketching problems are studied in Section \ref{sec_forward}. We will see that in the absence of noise we can choose $p=k$ for $k$-bandlimited signals to sketch estimates that are as good as least square estimates (Proposition \ref{prop_noiseless_forward}). In the presence of noise choosing $p>k$ helps in reducing noise (Proposition \ref{prop_noisy_forward}). The computation of optimal sampling matrices $\bbC$ and optimal estimation matrices $\bbH_s$ is intractable due to nonconvex objectives and binary constraints in the definition of the set $\ccalC$ in \eqref{eqn_sampling_matrix}. Heuristics for its solution are discussed in Section \ref{sec_heuristics}. As in the case of direct sketching, these heuristics still entail significant computation cost that is justifiable when solving a stream of estimation tasks. See Section \ref{sec_sims} for concrete examples.

%
\begin{remark}[Sketching and sampling] \label{rmk_sketching_and_sampling} \normalfont 
This paper studies sketching as a sampling problem to reduce the computational cost of computing the linear transformations of a vector $\bbx$. In its original definition sketching is not necessarily restricted to sampling, is concerned with the inverse problem only, and does not consider the joint design of sampling and estimation matrices \cite{mahoney11, woodruff14}. Sketching typically refers to the problem in \eqref{eqn_operator_inverse_sketching}-\eqref{eqn_operator_inverse_sketching_optimal_matrix} where the matrix $\bbC$ is not necessarily restricted to be a sampling matrix -- although it often is -- but any matrix such that the product $\bbC(\bbx+\bbw)$ can be computed with low cost. Our work differs in that we are trying to exploit a bandlimited model for the signal $\bbx$ to design optimal sampling matrices $\bbC$ along with, possibly, optimal computation matrices $\bbH_s$. Our work is also different in that we consider not only the inverse sketching problem of Section \ref{sssec_prob_inverse}  but also the direct sketching problem of Section \ref{sssec_prob_direct}.
\end{remark}

%
\section{Direct and inverse sketching as signal sampling} \label{sec_forward}


In this section we delve into the solutions of the direct and inverse sketching problems stated in Section \ref{sec_preliminaries}. We start  with the simple case where the observations are noise free (Section \ref{subsec_noisefree_forward}). This will be useful to gain insights on the solutions to the noisy formulations studied in Section \ref{subsec_noisy_forward}, and to establish links with the literature of sampling graph signals. Collectively, these results will also inform heuristic approaches to approximate the output of the linear transform in the direct sketching problem, and the least squares estimate in the inverse sketching formulation (Section \ref{sec_heuristics}). We consider operator sketching constraints in Section \ref{subsec_hsampling_forward}.

%
\subsection{Noise-free observations} \label{subsec_noisefree_forward}

Since in this noiseless scenario we have that $\bbw=\bbzero$ (cf. Figures~\ref{fig_forward}~and~\ref{fig_backward}), then the desired output for the direct sketching problem is $\bby=\bbH \bbx$ and the reduced-complexity approximation is given by $\hby=\bbH_s \bbC \bbx$. In the inverse problem we instead have $\bbx=\bbH^T\bby$, but assuming $\bbH$ is full rank then we can exactly invert the aforementioned relationship as $\bby:=\bbA_\mathrm{LS}\bbx=(\bbH \bbH^{T})^{-1}\bbH\bbx$. Accordingly, in the absence of noise we can equivalently view the inverse problem as a direct one whereby $\bbH=\bbA_\mathrm{LS}$. 

Next, we formalize the intuitive result that asserts that perfect estimation, namely that $\hby=\bby$, in the noiseless case is possible if $\bbx$ is a $k$-bandlimited signal [cf. \eqref{eqn_projection}] and the number of samples is $p\geq k$. To aid readability, the result is stated as a proposition.

%
\begin{proposition}\label{prop_noiseless_forward}
Let $\bbx \in \reals^{n}$ be a $k$-bandlimited signal and let $\bbH \in \reals^{m \times n}$ be a linear transformation. Let $\bbH_s \in \reals^{m \times p}$ be a reduced-input dimensionality sketch of $\bbH$, $p \le n$ and $\bbC \in \ccalC$ be a selection matrix. In the absence of noise ($\bbw=\mathbf{0}$), if $p=k$ and $\bbC^{*}$ is designed such that $\rank \{ \bbC^{*} \bbV_{k}\}=p=k$, then  $\hby=\bbH_s^{*} \bbC^{*} \bbx=\bby$ provided that the sketching matrix $\bbH_s^*$ is given by

	\begin{equation}\label{eqn_H3_noiseless_forward}
	\bbH_s^{*}=\left\{\begin{array}{cc}\bbH \bbV_{k} (\bbC^{*} \bbV_{k})^{-1},&\quad \textrm{Direct sketching},\\
	\bbA_\mathrm{LS} \bbV_{k} (\bbC^{*} \bbV_{k})^{-1},&\quad \textrm{Inverse sketching}.\end{array}\right.
	\end{equation}
\end{proposition}

\noindent The result follows immediately from e.g., the literature of sampling and reconstruction of bandlimited graph signals via selection sampling~\cite{chen15,tsitsvero16}. Indeed, if $\bbC^{*}$ is chosen such that $\rank \{ \bbC^{*} \bbV_{k}\}=p=k$,  then one can \emph{perfectly reconstruct} $\bbx$ from its samples $\bbx_s:=\bbC^{*}\bbx$ using the interpolation formula

\begin{equation}\label{eqn_reconstruct_x}
\bbx=\bbV_{k}(\bbC^{*} \bbV_{k})^{-1}\bbx_s.
\end{equation} 
The sketches $\bbH_s^{*}$ in \eqref{eqn_H3_noiseless_forward} follow after plugging \eqref{eqn_reconstruct_x} in $\bby=\bbH \bbx$ (or $\bby=\bbA_\mathrm{LS}\bbx$ for the inverse problem), and making the necessary identifications in $\hby=\bbH^{*}_s \bbC^{*} \bbx$. Notice that forming the inverse-mapping sketch $\bbH_{s}^{*}$ involves the (costly) computation of $(\bbH \bbH^{T})^{-1}$ within $\bbA_\mathrm{LS}$, but this is carried out entirely off line.	

In the absence of noise the design of $\bbC$ \textit{decouples} from that of $\bbH_s$. Towards designing $\bbC^{*}$, the $O(p^{3})$ complexity techniques proposed in~\cite{chen15} for finding a subset of $p$ rows of $\bbV_{k}$ that are linearly independent can be used here. Other existing methods to determine the most informative samples are relevant as well~\cite{Geert}. 

%
\subsection{Noisy observations} \label{subsec_noisy_forward}

Now consider the general setup described in Sections \ref{sssec_prob_direct} and \ref{sssec_prob_inverse}, where the noise vector signal $\bbw \in \reals^{n}$ is random and independent of $\bbx$, with $\mbE[\bbw]=\bbzero$, $\bbR_{w}=\mbE[\bbw \bbw^{T}] \in \reals^{n \times n}$ and $\bbR_{w} \succ \bbzero$. For the direct sketching formulation, we have $\bby = \bbH (\bbx+\bbw)$ and $\hby = \bbH_{s} \bbC (\bbx+\bbw)$ (see Figure \ref{fig_forward}). In the inverse problem we want to approximate the least squares estimate $\bbA_\mathrm{LS}(\bbx+\bbw)$ of $\bby$, with an estimate of the form $\hby = \bbH_{s} \bbC (\bbx+\bbw)$ as depicted in Figure \ref{fig_backward}.  Naturally, the joint design of $\bbH_{s}$ and $\bbC$ to minimize \eqref{eqn_direct_sketching_optimal_matrix} or \eqref{eqn_inverse_sketching_optimal_matrix} must account for the noise statistics. 

Said design will be addressed as a two-stage optimization that guarantees global optimality and proceeds in three steps. First, the optimal sketch $\bbH_{s}$ is expressed as a \textit{function} of $\bbC$. Second, such a function is substituted into the MMSE cost to yield a problem that depends only on $\bbC$. Third, the optimal $\bbH_s^*$ is found using the function in step one and the optimal value of $\bbC^*$ found in the step two. The result of this process is summarized in the following proposition; see the ensuing discussion and Appendix \ref{proof_prop_noisy_forward} for a short formal proof.

%
\begin{proposition} \label{prop_noisy_forward}
	Consider the direct and inverse sketching problems in the presence of noise [cf. \eqref{eqn_direct_sketching_optimal_matrix} or \eqref{eqn_inverse_sketching_optimal_matrix}]. Their solutions are $\bbC^*$ and $\bbH_s^{*} = \bbH_s^{*}(\bbC^{*})$, where
	
	\begin{equation}\label{eqn_H2_noisy_forward}
	\bbH_s^{*}(\bbC)=\left\{\begin{array}{cc}\bbH \bbR_{x} \bbC^{T} \left( \bbC (\bbR_{x}+\bbR_{w}) \bbC^{T} \right)^{-1},&\quad \textrm{Direct sketching},\\
	\bbA_\mathrm{LS} \bbR_{x} \bbC^{T} \left( \bbC ( \bbR_{x}+\bbR_{w}) \bbC^{T} \right)^{-1},&\quad \textrm{Inverse sketching}.\end{array}\right.
	\end{equation}
	For the direct sketching problem \eqref{eqn_direct_sketching_optimal_matrix}, the optimal sampling matrix $\bbC^*$ can be obtained as the solution to the problem
	
	\begin{equation}\label{eqn_prob1_forward}
	\min_{\bbC \in \ccalC} \ \tr \left[ \bbH \bbR_{x} \bbH^{T} - \bbH \bbR_{x} \bbC^{T}  \left( \bbC (\bbR_{x}+\bbR_{w}) \bbC^{T} \right)^{-1} \bbC \bbR_{x} \bbH^{T} \right]
	\end{equation}
	Likewise, for the inverse sketching problem \eqref{eqn_inverse_sketching_optimal_matrix}, $\bbC^*$ is the solution of
	
	\begin{equation} \label{eqn_prob1_backward_x}
	\min_{\bbC \in \ccalC} \ \tr \big[ \bbR_{x} - \bbH^{T}\bbA_{\mathrm{LS}} \bbR_{x} \bbC^{T} \left( \bbC (\bbR_{x}+\bbR_{w}) \bbC^{T} \right)^{-1} \bbC \bbR_{x} \bbA_{\mathrm{LS}}^T\bbH \big].
	\end{equation}
\end{proposition}

\noindent For the sake of argument, consider now the direct sketching problem. The optimal sketch $\bbH_s^{*}$ in \eqref{eqn_H2_noisy_forward} is tantamount to $\bbH$ acting on a preprocessed version of $\bbx$, using the matrix $\bbR_{x} \bbC^{T} ( \bbC (\bbR_{x}+\bbR_{w}) \bbC^{T} )^{-1}$. What this precoding entails is, essentially, choosing the samples of $\bbx$ with the optimal tradeoff between the signal in the factor $\bbR_{x} \bbC^{T}$ and the noise in the inverse term $( \bbC (\bbR_{x}+\bbR_{w}) \bbC^{T} )^{-1}$.  This is also natural from elementary results in linear MMSE theory~\cite{kay93-estimation}. Specifically, consider forming the MMSE estimate of $\bbx$ given observations $\bbx_s=\bbC\bbx_s+\bbw_s$, where $\bbw_s:=\bbC\bbw$ is a zero-mean sampled noise with covariance matrix $\bbR_{w_s}=\bbC\bbR_w\bbC^T$. Said estimator is given by 

\begin{equation}\label{eqn_MMSE_x}
\hbx = \bbR_{x} \bbC^{T} ( \bbC (\bbR_{x}+\bbR_{w}) \bbC^{T} )^{-1}\bbx_s.
\end{equation}
Because linear MMSE estimators are preserved through linear transformations such as $\bby=\bbH\bbx$, then the sought MMSE estimator of the response signal is $\hby=\bbH\hbx$. Hence, the expression for $\bbH_s^{*}$ in \eqref{eqn_H2_noisy_forward} follows. Once more, the same argument holds for the inverse sketching problem after replacing $\bbH$ with the least squares operator $\bbA_\mathrm{LS}=(\bbH\bbH^{T})^{-1}\bbH$. Moreover, note that \eqref{eqn_prob1_backward_x} stems from minimizing $\mbE[\| \bbx - \bbH^{T} \hby\|_{2}^{2}]$ as formulated in \eqref{eqn_inverse_sketching_optimal_matrix}, which is the standard objective function in linear regression problems. Minimizing $\mbE[\|\bby - \hby\|_{2}^{2}]$ for the inverse problem is also a possibility, and one obtains solutions that closely resemble the direct problem. Here we opt for the former least squares estimator defined in \eqref{eqn_inverse_sketching_optimal_matrix}, since it is the one considered in the sketching literature \cite{woodruff14}; see Section \ref{sec_sims} for extensive performance comparisons against this baseline.

Proposition~\ref{prop_noisy_forward} confirms that the optimal selection matrix $\bbC^{*}$ is obtained by solving  \eqref{eqn_prob1_forward} which, after leveraging the expression in \eqref{eqn_H2_noisy_forward}, only requires knowledge of the given matrices $\bbH$, $\bbR_{x}$ and $\bbR_{w}$. The optimal sketch is then found substituting $\bbC^{*}$ into \eqref{eqn_H2_noisy_forward} as $\bbH_{s}^{*}=\bbH_{s}^{*}(\bbC^{*})$, a step incurring $O(mnp+p^3)$ complexity. Naturally, this two-step solution procedure resulting in $\{\bbC^{\ast},\bbH_{s}^{\ast}(\bbC^{\ast})\}$ entails no loss of optimality~\cite[Section~4.1.3]{bova04}, while effectively reducing the dimensionality of the optimization problem. Instead of solving a problem with $p(m+n)$ variables [cf. \eqref{eqn_direct_sketching_optimal_matrix}], we first solve a problem with $pn$ variables [cf. \eqref{eqn_prob1_forward}] and, then, use the closed-form in \eqref{eqn_H2_noisy_forward} for the remaining $pm$ unknowns. The practicality of the approach relies on having a closed-form for $\bbH_s^*(\bbC)$. While this is possible for the quadratic cost in \eqref{eqn_direct_sketching_optimal_matrix}, it can be challenging for other error metrics or nonlinear signal models. In those cases, schemes such as alternating minimization, which solves a sequence of problems of size $pm$ (finding the optimal $\bbH_s$ given the previous $\bbC$) and $pn$ (finding the optimal $\bbC$ given the previous $\bbH_s$), can be a feasible way to bypass the (higher dimensional and non-convex) joint optimization. 

The next proposition establishes that \eqref{eqn_prob1_forward} and \eqref{eqn_prob1_backward_x}, which yield the optimal $\bbC^{*}$, are equivalent to \emph{binary} optimization problems with a linear objective function and subject to linear matrix inequality (LMI) constraints. 

%
\begin{proposition} \label{prop_noisy_forward_sdp}
Let $\bbc \in \{0,1\}^{n}$ be the binary vector that contains the diagonal elements of $\bbC^{T}\bbC$, i.e. $\bbC^{T} \bbC=\diag(\bbc)$. Then, in the context of Proposition~\ref{prop_noisy_forward}, the optimization problem \eqref{eqn_prob1_forward} over $\bbC$ is equivalent to

\begin{align}\label{eqn_prob2_forward}
	\min_{\substack{\bbc \in \{0,1\}^{n},\\\bbY, \barbC_{\alpha}}}\
	& \tr \left[ \bbY \right] \\
	\st\
	& \barbC_{\alpha}=\alpha^{-1}\diag(\bbc) \ , \
	\bbc^{T} \bbone_{n}=p \nonumber \\
	& \begin{bmatrix} \bbY-\bbH \bbR_{x} \bbH^{T} + \bbH \bbR_{x} \barbC_{\alpha} \bbR_{x} \bbH^{T} & \bbH \bbR_{x} \barbC_{\alpha} \\ \barbC_{\alpha} \bbR_{x} \bbH^{T} & \barbR_{\alpha}^{-1} + \barbC_{\alpha} \end{bmatrix} \succeq \bbzero \nonumber
\end{align}
where $\barbR_{\alpha}=(\bbR_{x}+\bbR_{w}-\alpha \bbI_{n})$, $\barbC_{\alpha}$ and $\bbY\in \reals^{m\times m}$ are auxiliary variables and $\alpha>0$ is any scalar satisfying $\barbR_{\alpha}\succ \bbzero$.

Similarly, \eqref{eqn_prob1_backward_x} is equivalent to a problem identical to \eqref{eqn_prob2_forward} except for the LMI constraint that should be replaced with

\begin{equation}
\begin{bmatrix} \bbY - \bbR_{x} + \bbH^{T}\bbA_{\mathrm{LS}} \bbR_{x} \barbC_{\alpha}\bbR_{x} \bbA_{\mathrm{LS}}^T\bbH & \bbH^{T}\bbA_{\mathrm{LS}} \bbR_{x} \barbC_{\alpha} \\ \barbC_{\alpha} \bbR_{x} \bbA_{\mathrm{LS}}^T\bbH & \barbR_{\alpha}^{-1}+\barbC_{\alpha} \end{bmatrix} \succeq \bbzero.\label{eqn_prob2_backward_x}
\end{equation}
\end{proposition}

\noindent See Appendix \ref{proof_prop_noisy_forward_sdp} for a proof. Problem \eqref{eqn_prob2_forward} is an SDP optimization modulo the binary constraints on vector $\bbc$, which can be relaxed to yield a convex optimization problem (see Remark~\ref{rmk_numerical_instabilities} for comments on the value of $\alpha$). Once the relaxed problem is solved, the solution can be binarized again to recover a feasible solution $\bbC\in\ccalC$. This convex relaxation procedure is detailed in Section~\ref{sec_heuristics}. 

%
\begin{remark}[Numerical considerations]\label{rmk_numerical_instabilities}\normalfont
While there always exists $\alpha>0$ such that $\barbR_{\alpha}=(\bbR_{x}+\bbR_{w}\blue{-\alpha \bbI_{n}})$ is invertible, this value of $\alpha$ has to be smaller than the smallest eigenvalue of $\bbR_{x}+\bbR_{w}$. In low-noise scenarios this worsens the condition number of $\barbR_{\alpha}$, creating numerical instabilities when inverting said matrix (especially for large graphs). Alternative heuristic solutions to problems \eqref{eqn_prob1_forward} and \eqref{eqn_prob2_forward} (and their counterparts for the inverse sketching formulation) are provided in Section~\ref{ssec_conv_relax}.
\end{remark}

\subsection{Operator sketching} \label{subsec_hsampling_forward}

As explained in Sections \ref{sssec_prob_direct} and \ref{sssec_prob_inverse}, there may be setups where it is costly (or even infeasible) to freely design the new operator $\bbH_s$ with entries that do not resemble those in $\bbH$. This can be the case in distributed setups where the values of $\bbH$ cannot be adapted, or when the calculation (or storage) of the optimal $\bbH_s$ is impossible. An alternative to overcome this challenge consists in forming the sketch by sampling $p$ \textit{columns} of $\bbH$, i.e., setting $\bbH_s=\bbH\bbC^{T}$ and optimizing for $\bbC$ in the sense of \eqref{eqn_operator_direct_sketching_optimal_matrix} or \eqref{eqn_operator_inverse_sketching_optimal_matrix}. Although from an MSE performance point of view such an \textit{operator sketching} design is suboptimal [cf. \eqref{eqn_H2_noisy_forward}], numerical tests carried out in Section \ref{sec_sims} suggest it can sometimes yield competitive performance. The optimal \textit{sampling} strategy for sketches within this restricted class is given in the following proposition; see Appendix \ref{proof_prop_hsampling_forward} for a sketch of the proof.

%
\begin{proposition} \label{prop_hsampling_forward}
	Let $\bbH_s=\bbH \bbC^{T}$ be constructed from a subset of $p$ columns of $\bbH$. Then, the optimal sampling matrix $\bbC^{*}$ defined in  \eqref{eqn_operator_direct_sketching_optimal_matrix} can be recovered from the diagonal elements $\bbc^{*}$ of $(\bbC^{*})^{T} \bbC^*=\diag(\bbc^{*})$, where $\bbc^{*}$ is the solution to the following problem
	
	\begin{align}
	\min_{\substack{\bbc \in \{0,1\}^{n},\\\bbY, \barbC}}\
	& \tr [\bbY] \label{eqn_prob2_hsampling_forward} \\
	\st\
	& \barbC=\diag(\bbc) \ , \ \bbc^{T} \bbone_{n} = p \nonumber \\
	& \begin{bmatrix} \bbY - \bbH \bbR_{x} \bbH^{T} + 2 \bbH \barbC \bbR_{x} \bbH^{T} & \bbH \barbC \\ \barbC \bbH^{T} & (\bbR_{x} + \bbR_{w})^{-1} \end{bmatrix} \succeq \bbzero.
	\nonumber
	\end{align}
Likewise,  the optimal sampling matrix $\bbC^{*}$ defined in \eqref{eqn_operator_inverse_sketching_optimal_matrix} can be recovered from the solution to a  problem identical to \eqref{eqn_prob2_hsampling_forward} except for the LMI constraint that should be replaced with

\begin{equation}\label{eqn_prob2_hsampling_backward_x}
\begin{bmatrix} \bbY - \bbR_{x} + 2 \bbH^{T} \bbH \barbC \bbR_{x} & \bbH^{T} \bbH \barbC \\ \barbC \bbH^{T} \bbH & (\bbR_{x}+\bbR_{w})^{-1} \end{bmatrix} \succeq \bbzero.
\end{equation}

\end{proposition}

In closing, we reiterate  that solving the optimization problems stated in Propositions \ref{prop_noisy_forward_sdp} and \ref{prop_hsampling_forward} is challenging because of the binary decision variables $\bbc \in \{0,1\}^{n}$. Heuristics for approximate solution with manageable computational cost are discussed next.

%


%
\section{Heuristic approaches} \label{sec_heuristics}


In this section, several heuristics are outlined for tackling the linear sketching problems described so far. The rationale is that oftentimes the problems posed in Section~\ref{sec_forward} can be intractable, ill-conditioned, or, just too computationally expensive even if carried out off line. In fact, the optimal solution $\bbC^{\ast}$ to \eqref{eqn_prob1_forward} or \eqref{eqn_prob1_backward_x} can be obtained by evaluating the objective function in each one of the $\binom{n}{p}$ possible solutions. Table \ref{table_computational-complexity} lists the complexity of each of the proposed methods. Additionally, the time (in seconds) taken to run the simulation related to Figure~\ref{fig_digits} is also included in the table for comparison. In all cases, after obtaining $\bbC^{\ast}$, forming the optimal value of $\bbH_{s}^{\ast}$ in \eqref{eqn_H2_noisy_forward} entails $O(mnp+p^3)$ operations.

%
\subsection{Convex relaxation (SDP)} \label{ssec_conv_relax}

Recall that the main difficulty when solving the optimization problems in Propositions~\ref{prop_noisy_forward_sdp}~and~\ref{prop_hsampling_forward} are the \emph{binary} constraints that render the problems non-convex and, in fact, NP-hard. A {standard} alternative to overcome this difficulty is to relax the binary constraint $\bbc\in \{0,1\}^{n}$ on the sampling vector as $\bbc \in [0,1]^{n}$. This way, the optimization problem \eqref{eqn_prob2_forward}, or alternatively, with LMI constraints \eqref{eqn_prob2_backward_x}, becomes convex and can be solved with polynomial complexity in $O((m+n)^{3.5})$ operations as per the resulting SDP formulation  \cite{bova04}.

Once a solution to the relaxed problem is obtained, two ways of recovering a binary vector $\bbc$ are considered. The first one consists in computing $\bbp_{c}=\bbc/\|\bbc\|_{1}$, which can be viewed as a probability distribution over the samples (SDP-Random). These samples are then drawn at random from this distribution; see~\cite{puy15}. This should be done once, off line, and the same selection matrix used for every incoming input (or output). The second one is a deterministic method referred to as thresholding (SDP-Thresh.), which simply consists in setting the largest $p$ elements to $1$ and the rest to $0$. Since the elements in $\bbc$ are non-negative, note that the constraint $\bbc^{T} \bbone_{n} = p$ considered in the optimal sampling formulation can be equivalently rewritten as $\|\bbc\|_1 = p$. Using a dual approach, this implies that the objective of the optimization problem is implicitly augmented with an $\ell_1$-norm penalty $\lambda\|\bbc\|_1$, whose regularization parameter $\lambda$ corresponds to the associated Lagrange multiplier. In other words, the formulation is implicitly promoting sparse solutions. The adopted thresholding is a natural way to approximate the sparsest $\ell_0$-(pseudo) norm solution with its convex $\ell_1$-norm surrogate. An alternative formulation of the convex optimization problem can thus be obtained by replacing the constraint $\bbc^{T} \bbone_{n} = p$ with a penalty $\lambda \| \bbc \|_{1}$ added to the objective, with a hyperparameter $\lambda$. While this approach is popular, for the simulations carried out in Section~\ref{sec_sims} we opted to keep the constraint $\bbc^{T} \bbone_{n} = p$ to have explicit control over the number of selected samples $p$, and to avoid tuning an extra hyperparameter $\lambda$.

\begin{table}[tb]
	\centering
	\begin{tabular}{|l|c|c|} \hline
				&			&	   \\
	Method			& Number of operations & Time [s]  \\
				&			& \\ \hline
	 & 	&  \\
	Optimal solution	& $\textstyle O \left(\! \left(\!\!\begin{array}{c} n\\p \end{array}\!\!\right)\!\right)$ 	& $>10^{6}$ \\ 
	 & 	& \\
	Convex relaxation (SDP)		& $\displaystyle O \left((m+n)^{3.5}\right)$ 	& $25.29$\\
	 & 	& \\
	Noise-blind heuristic	& $\displaystyle O \left(n \log n+nm\right)$ 		& $0.46$ \\
	 & 	& \\
	 Noise-aware heuristic	& $\displaystyle O \left(n^{3}+nm\right)$ 		& $21.88$ \\
	 & 	& \\
	Greedy approach		& $\displaystyle O \left(mn^{2}p^{2}+np^{4}\right)$ 		& $174.38$\\
	 & 	& \\ \hline
	\end{tabular}
	\caption{\small Computational complexity of the heuristic methods proposed in Section \ref{sec_heuristics} to solve optimization problems in Section \ref{sec_forward}: (i) Number of operations required; (ii) Time (in seconds) required to solve the problem in Section~\ref{subsec_digits}, see Figure~\ref{fig_digits}. Also, for the sake of comparison, we have included a row containing the cost of the optimal solution, meaning the cost of solving the optimization problems exactly with no relaxation on the binary constraints.}
	\label{table_computational-complexity}
\end{table}


%
\subsection{Noise-aware heuristic (NAH)} \label{ssec_noiseaware}
  
A heuristic that is less computationally costly than the SDP in Section~\ref{ssec_conv_relax} can be obtained as follows. Consider for instance the objective function of the direct problem \eqref{eqn_prob1_forward}

\begin{equation} 
    \tr \left[ \bbH \bbR_{x} \bbH^{T} - \bbH \bbR_{x} \bbC^{T} \left( \bbC (\bbR_{x}+\bbR_{w}) \bbC^{T} \right)^{-1} \bbC \bbR_{x} \bbH^{T} \right]
\end{equation}
where we note that the noise imposes a tradeoff in the selection of samples. More precisely, while some samples are very important in contributing to the transformed signal, as determined by the rows of $\bbR_{x} \bbH^{T}$, those same samples might provide very noisy measurements, as determined by the corresponding submatrix of $(\bbR_{x}+\bbR_{w})$. Taking this tradeoff into account, the proposed noise-aware heuristic (NAH) consists of selecting the rows of $(\bbR_{x}+\bbR_{w})^{-1/2} \bbR_{x} \bbH^{T}$ with highest $\ell_{2}$ norm. This polynomial-time heuristic entails $O(n^{3})$ operations to compute the inverse square root of $(\bbR_{x}+\bbR_{w})$ \cite{frommer09} and $O(mn)$ to compute its multiplication with $\bbR_{x}\bbH^{T}$ as well as the norm. Note that $(\bbR_{x}+\bbR_{w})^{-1/2}\bbR_{x}\bbH^{T}$ resembles a signal-to-noise ratio (SNR), and thus the NAH is attempting to maximize this measure of SNR.

%
\subsection{Noise-blind heuristic (NBH)} \label{ssec_noiseblind}

Another heuristic that incurs an even lower computational cost can also be obtained by inspection of \eqref{eqn_prob1_forward}. Recall that the complexity of the NAH is dominated by the computation of  $(\bbR_{x}+\bbR_{w})^{-1/2}$. An even faster heuristic solution can thus be obtained by simply ignoring this term, which accounted for the noise present in the chosen samples. Accordingly, in what we termed the noise-blind heuristic (NBH) we simply select the $p$ rows of $\bbR_{x}\bbH^{T}$ that have maximum $\ell_{2}$ norm. The resulting NBH is straightforward to implement, entails $O(mn)$ operations for computing the norm of $\bbR_{x}\bbH^{T}$ and $O(n \log n)$ operations for the sorting algorithm \cite{knuth98}. It is shown in Section \ref{sec_sims} to yield satisfactory performance, especially if the noise variance is low or the linear transform has favorable structure. In summary, the term noise-blind stems from the fact that we are selecting the samples that yield the highest output energy as measured by $\bbR_{x}\bbH^{T}$, while being completely agnostic to the noise corrupting those same samples.

The analysis of both the NAH (Section~\ref{ssec_noiseaware}) and NBH (Section~\ref{ssec_noiseblind}) can be readily extended to the linear inverse problem by inspecting the objective function in \eqref{eqn_prob1_backward_x}.

%
\subsection{Greedy approach} \label{ssec_greedy}

Another alternative to approximate the solution {of} \eqref{eqn_prob1_forward} and \eqref{eqn_prob1_backward_x} over $\ccalC$, is to implement an iterative greedy algorithm that adds samples to the sampling set incrementally. At each iteration, the sample that reduces the MSE the most is incorporated to the sampling set. Considering problem \eqref{eqn_prob1_forward} as an example, first, one-by-one all $n$ samples are tested and the one that yields the lowest value of the objective function in \eqref{eqn_prob1_forward} is added to the sampling set. Then, the sampling set is augmented with one more sample by choosing the one that yields the lowest optimal objective among the remaining $n-1$ ones. The procedure is repeated until $p$ samples are selected in the sampling set. This way only $n + (n-1) + \cdots + (n-(p-1))<np$ evaluations of the objective function \eqref{eqn_prob1_forward} are required. Note that each evaluation of \eqref{eqn_prob1_forward} entails $O(mnp+p^{3})$ operations, so that the overall cost of the greedy approach is $O(mn^{2}p^{2}+np^{4})$. Greedy algorithms have well-documented merits for sample selection, even for non-submodular objectives like the one in \eqref{eqn_prob1_forward}; see \cite{chepuri16,chamon16}.

%
\section{Numerical Examples} \label{sec_sims}


To demonstrate the effectiveness of the sketching methods developed in this paper, five numerical test cases are considered. In Sections \ref{subsec_gft_sims} and \ref{subsec_big_data} we look at the case where the linear transform to approximate is a graph Fourier transform (GFT) and the signals are bandlimited on a given graph~\cite{shuman13,sandryhaila14}. Then, in Section~\ref{subsec_sensor} we consider an (inverse) linear estimation problem in a wireless sensor network. The transform to approximate is the (fat) linear estimator and choosing samples in this case boils down to selecting the sensors acquiring the measurements \cite[Section VI-A]{Geert}. For the fourth test, in Section \ref{subsec_digits}, we look at the classification of digits from the MNIST Handwritten Digit database \cite{mnist}. By means of principal component analysis (PCA), we can accurately describe these images using a few coefficients, implying that they (approximately) belong to a lower dimensional subspace given by a few columns of the covariance matrix. Finally, in Section~\ref{subsec_author} we consider the problem of authorship attribution to determine whether a given text belongs to some specific author or not, based on the stylometric signatures dictated by  word adjacency networks \cite{Segarra15-WANs}.

Throughout, we compare the performance in approximating the output of the linear transform of the sketching-as-sampling technique presented in this paper (implemented by each of the five heuristics introduced) and compare it to that of existing alternatives. More specifically, we consider:

\begin{enumerate}[a)]
	\item The sketching-and-sampling algorithms proposed in this paper, namely: a1) the random sampling scheme based on the convex relaxation (SDP-Random, Section \ref{ssec_conv_relax}); a2) the thresholding sampling schemed based on the convex relaxation (SDP-Thresh., Section \ref{ssec_conv_relax}); a3) the noise-aware heuristic (NAH, Section \ref{ssec_noiseaware}); a4) the noise-blind heuristic (NBH, Section \ref{ssec_noiseblind}); a5) the greedy approach (Section \ref{ssec_greedy}); and a6) the operator sketching methods that directly sample the linear transform (SLT) [cf. Section~\ref{subsec_hsampling_forward}] by solving the problems in Proposition~\ref{prop_hsampling_forward} using the methods a1)-a5).
	
	\item  Algorithms for sampling bandlimited signals [cf. \eqref{eqn_projection}], namely:  b1) the experimental design sampling (EDS) method proposed in \cite{varma15}; and b2) the spectral proxies (SP) greedy-based method proposed in \cite[Algorithm 1]{anis16}. In particular, we note that the EDS method in \cite{varma15} computes a distribution where the probability of selecting the $i$th element of $\bbx$ is proportional to the norm of the $i$th row of the matrix $\bbV$ that determines the subspace basis. This gives rise to three different EDS methods depending on the norm used: EDS-1 when using the $\ell_{1}$ norm, EDS-2 when using the $\ell_{2}$ norm \cite{puy15} and EDS-$\infty$ when using the $\ell_{\infty}$ norm \cite{varma15}.
	
	\item Traditional sketching algorithms, namely Algorithms 2.4, 2.6 and 2.11 described in (the tutorial paper) \cite{woodruff14}, and respectively denoted in the figures as Sketching 2.4, Sketching 2.6 and Sketching 2.11. Note that these schemes entail different (random) designs of the matrix $\bbC$ in \eqref{eqn_inverse_sketching} that do not necessarily entail sampling (see Remark~\ref{rmk_sketching_and_sampling}). They can only be used for the inverse problem  \eqref{eqn_inverse_sketching_optimal_matrix}, hence they will be tested only in Sections~\ref{subsec_gft_sims}~and~\ref{subsec_sensor}.
	
	\item The optimal linear transform $\hby = \bbA^{\ast} (\bbx + \bbw)$ that minimizes the MSE, operating on the entire signal, without any sampling nor sketching. Exploiting the fact that the noise $\bbw$ and the signal $\bbx$ are uncorrelated, these estimators are $\bbA^{\ast} = \bbH$ for the direct case and $\bbA^{\ast} =\bbA_\mathrm{LS}= (\bbH \bbH^{T})^{-1} \bbH$ for the inverse case. We use these method as the corresponding baselines (denoted as Full). Likewise, for the classification problems in Sections~\ref{subsec_digits}~and~\ref{subsec_author}, the baseline is the corresponding classifier operating on the entire signal, without any sampling (denoted as SVM).
\end{enumerate}

To aid readability and reduce the number of curves on the figures presented next, only the best-performing among the three EDS methods in b1) and the best-performing of the three sketching algorithms in c) are shown.

%
\subsection{Approximating the GFT as an inverse problem} \label{subsec_gft_sims}

Obtaining alternative data representations that would offer better insights and facilitate the resolution of specific tasks is one of the main concerns in signal processing and machine learning. In the particular context of GSP, a $k$-bandlimited graph signal $\bbx = \bbV_{k} \tbx_{k}$ can be described as belonging to the $k$-dimensional subspace spanned by the $k$ eigenvectors $\bbV_{k}= [\bbv_{1},\ldots,\bbv_{k}] \in \mbC^{k \times n}$ of the graph shift operator $\bbS$ [cf. \eqref{eqn_projection}, \eqref{eqn_recovery}]. The coefficients $\tbx_{k} \in \mbC^{k}$ are known as the GFT coefficients and offer an alternative representation of $\bbx$ that gives insight into the modes of variability of the graph signal with respect to the underlying graph topology \cite{sandryhaila14}.

Computing the GFT coefficients of a bandlimited signal following $\bbx = \bbV_{k} \tbx_{k}$ can be modeled as an inverse problem where we have observations of the output $\bbx$, which are a transformation of $\bby = \tbx_{k}$ through a linear operation $\bbH^{T} = \bbV_{k}$ (see Section~\ref{sssec_prob_inverse}). We can thus reduce the complexity of computing the GFT of a sequence of graph signals, by adequately designing a sampling pattern $\bbC$ and sketching matrix $\bbH_{s}$ that operate only on $p \ll n$ samples of each graph signal $\bbx$, instead of solving $\bbx = \bbV_{k} \tbx_{k}$ for the entire graph signal $\bbx$ (cf. Figure~\ref{fig_backward}). This reduces the computational complexity by a factor of $n/p$, serving as a fast method of obtaining the GFT coefficients.

In what follows, we set the graph shift operator $\bbS$ to be the adjacency matrix of the underlying undirected graph $\ccalG$. Because the shift is symmetric, it can be decomposed as $\bbS = \bbV \bbLambda \bbV^{T}$. So the GFT to approximate is $\bbV^{T}$, a real orthonormal matrix that projects signals onto the eigenvector space of the adjacency of $\ccalG$. With this notation in place, for this first experiment we assume to have access to $100$ noisy realizations of this $k$-bandlimited graph signal $\{\bbx_{t}+\bbw_{t}\}_{t=1}^{100}$, where $\bbw_{t}$ is zero-mean Gaussian noise with $\bbR_{w} = \sigma_{w}^{2} \bbI_{n}$. The objective is to compute $\{\tbx_{k,t}\}_{t=1}^{100}$, that is, the $k$ active GFT coefficients of each one of these graph signals. 

To run the algorithms, we consider two types of undirected graphs: a stochastic block model (SBM) and a small-world (SW) network. Let $\ccalG_{\textrm{SBM}}$ denote a SBM network with $n$ total nodes and $c$ communities with $n_{b}$ nodes in each community, $b=1,\ldots,c$, $\sum_{b=1}^{c} n_{b} = n$ \cite{decelle11}. The probability of drawing an edge between nodes within the same community is $p_{b}$ and the probability of drawing an edge between any node in community $b$ and any node in community $b'$ is $p_{bb'}$. Similarly, $\ccalG_{\textrm{SW}}$ describes a SW network with $n$ nodes characterized by parameters $p_{e}$ (probability of drawing an edge between nodes) and $p_r$ (probability of rewiring edges) \cite{watts98}. In the first experiment, we study signals $\bbx \in \reals^{n}$ supported on either the SBM or the SW networks. In each of the test cases presented, $\ccalG \in \{\ccalG_{\textrm{SBM}}, \ccalG_{\textrm{SW}}\}$ denotes the underlying graph, and $\bbA \in \{0,1\}^{n \times n}$ its associated adjacency matrix. The simulation parameters are set as follows. The number of nodes is $n=96$ and the bandwidth is $k=10$. For the SBM, we set $c=4$ communities of $n_{b}=24$ nodes in each, with edge probabilities $p_{b}=0.8$ and $p_{bb'}=0.2$ for  $b\neq b'$. For the SW case, we set the edge and rewiring probabilities as $p_{e} = 0.2$ and $p_{r}=0.7$. The metric to assess the reconstruction performance is  the relative mean squared error (MSE) computed as $\mbE[\|\hby - \tbx_{k}\|_{2}^{2}]/\mbE[\|\tbx_{k}\|_{2}^{2}]$. We estimate the MSE by simulating $100$ different sequences of length $100$, totaling $10,000$ signals. Each of these signals is obtained by simulating $k=10$ i.i.d. zero-mean, unit-variance Gaussian random variables, squaring them to obtain the GFT $\tbx_{k}$, and finally computing $\bbx = \bbV_{k} \tbx_{k}$. We repeat this simulation for $5$ different random graph realizations. For the methods that use the covariance matrix $\bbR_{x}$ as input, we estimate $\bbR_{x}$ from $500$ realizations of $\bbx$, which we regarded as training samples and are not used for estimating the relative MSE. Finally, for the methods in which the sampling is random, i.e., a1), b1), and c), we perform the node selection $10$ different times and average the results.

\begin{figure}[t]
    \captionsetup[subfigure]{justification=centering}
    \centering
    
    \begin{subfigure}{0.49\textwidth}
        \centering
        \includegraphics[width=\threeplotsize\textwidth]{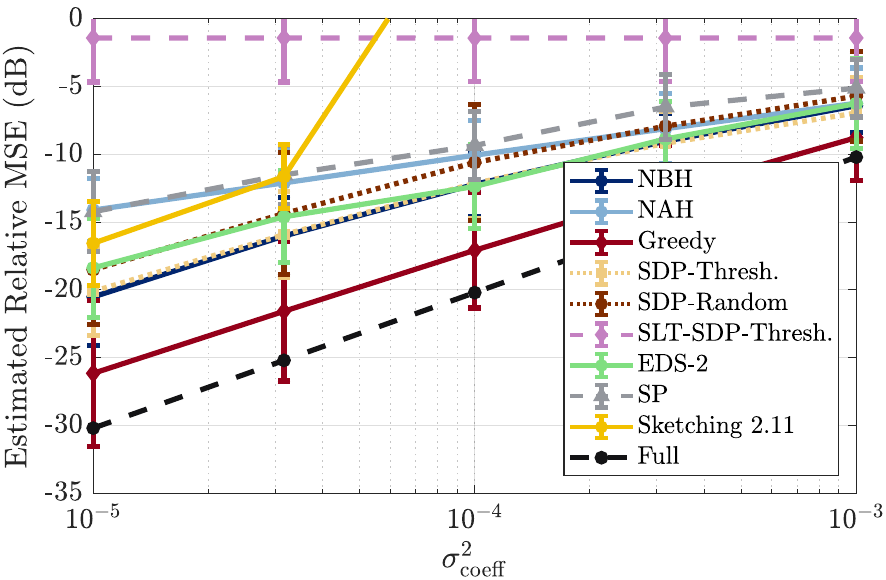}
        \caption{Stochastic block model}
        \label{SBM-noise}
    \end{subfigure}
    \hfill
    \begin{subfigure}{0.49\textwidth}
        \centering
        \includegraphics[width=\threeplotsize\textwidth]{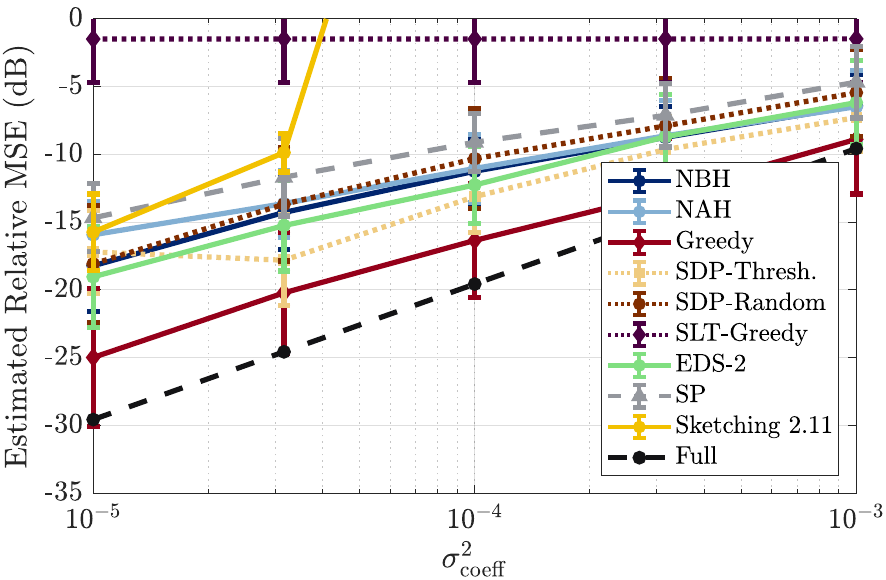}
        \caption{Small world}
        \label{SW-noise}
    \end{subfigure}
    \caption{Approximating the GFT as an inverse sketching problem. Relative estimated MSE as a function of noise. Legends: SDP-Random (scheme in a1); SDP-Thresh. (scheme in a2); NAH (scheme in a3), NBH (scheme in a4); Greedy (scheme in a5); SLT, EDS-1, EDS-2 (schemes in b1), SP (scheme in b2); Sketching 2.6, Sketching 2.11 (schemes in c); and baseline using the full signal (no sampling).}
    \label{fig_inverse_100_10_noise}
\end{figure}

\begin{figure}[t]    
    \begin{subfigure}{0.49\textwidth}
        \centering
        \includegraphics[width=\threeplotsize\textwidth]{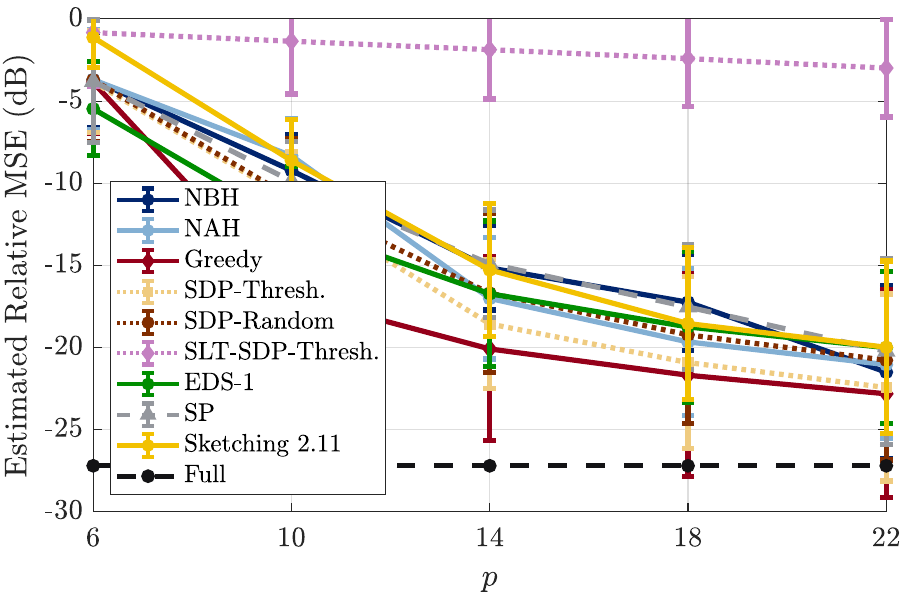}
        \caption{Stochastic block model}
        \label{SBM-p} 
    \end{subfigure}
    \hfill
    \begin{subfigure}{0.49\textwidth}
        \centering
        \includegraphics[width=\threeplotsize\textwidth]{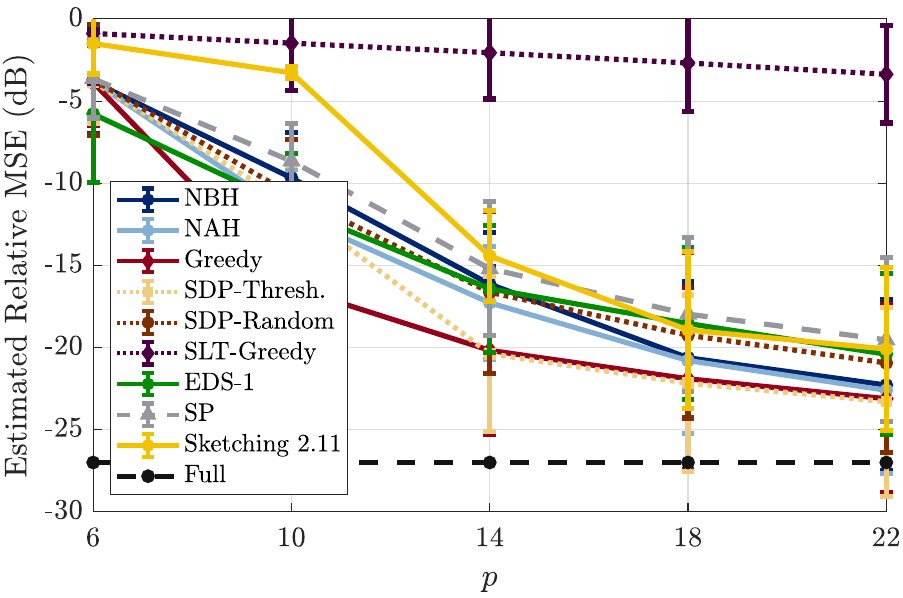}
        \caption{Small world}
        \label{SW-p} 
    \end{subfigure}
    \caption{Approximating the GFT as an inverse sketching problem. Relative estimated MSE as a function of the number of samples used. Legends: SDP-Random (scheme in a1); SDP-Thresh. (scheme in a2); NAH (scheme in a3), NBH (scheme in a4); Greedy (scheme in a5); SLT, EDS-1, EDS-2 (schemes in b1), SP (scheme in b2); Sketching 2.6, Sketching 2.11 (schemes in c); optimal solution using the full signal.}
    \label{fig_inverse_100_10_p}
\end{figure}

For each of the graph types (SBM and SW), we carry out two parametric simulations. In the first one, we consider the number of selected samples to be fixed to $p=k=10$ and consider different noise power levels  $\sigma^{2}_{w} = \sigma^{2}_{\textrm{coeff}} \cdot \mbE[\|\bbx\|^{2}]$ by varying $\sigma^{2}_{\textrm{coeff}}$ from $10^{-5}$ to $10^{-3}$ and where $\mbE[\|\bbx\|^{2}]$ is estimated from the training samples, see Figure~\ref{fig_inverse_100_10_noise} for results. For the second simulation, we fix the noise to $\sigma^{2}_{\textrm{coeff}}=10^{-4}$ and vary the number of selected samples $p$ from $6$ to $22$, see Figure~\ref{fig_inverse_100_10_p}.

%

First, Figures~\ref{SBM-noise}~and~\ref{SW-noise} show the estimated relative MSE as a function of $\sigma_{\textrm{coeff}}^{2}$ for fixed $p=k=10$ for the SBM and the SW graph supports, respectively. We note that, for both graph supports, the greedy approach in a5) outperforms all other methods and is, at most, $5\text{dB}$ worse than the baseline which computes the GFT using the full signal, while saving $10$ times computational cost in the online stage. Then, we observe that the SDP-Thresh. method in a2) is the second best method, followed closely by the EDS-2 in b1). We observe that both NAH and NBH heuristics of a3) and a4) have similar performance, with the NBH working considerably better in the low-noise scenario. This is likely due to the suboptimal account of the noise carried out by the NAH (see Section~\ref{ssec_noiseaware}). NBH works as well as the SDP-Thresh. method in the SBM case. With respect to off-line computational complexity, we note that the EDS-2 method incurs in an off-line cost of $O(n^{3}+n^{2}+n \log(n))$ which, in this problem, reduces to $O(10^{6})$, while the greedy scheme has a cost of $O(10^{7})$; however, the greedy approach performs almost one order of magnitude better in the low-noise scenario. Alternatively, the NBH heuristic has a comparable performance to the EDS-2 on both graph supports, but incurs in an off-line cost of $O(10^{3})$.

%

Second, the relative MSE as a function of the number of samples $p$ for fixed noise $\sigma_{\textrm{coeff}}^{2}$ for the SBM and SW networks is plotted in Figures~\ref{SBM-p}~and~\ref{SW-p}, respectively. In this case, we note that the greedy approach of a5) outperforms all other methods, but its relative gain in terms of MSE is inferior. This would suggest that less computationally expensive methods like the NBH are preferable, especially for higher number of samples. Additionally, we observe that for $p < k$, the EDS graph signal sampling technique works better. For $p=22$ selected nodes, the best performing sketching-as-sampling technique (the greedy approach) performs $5\text{dB}$ worse than the baseline, but incurring in only $23\%$ of the online computational cost.

\FloatBarrier

%
\subsection{Approximating the GFT in a large-scale network} \label{subsec_big_data}

The GFT coefficients can also be computed via a matrix multiplication $\tbx_{k} = \bbV_{k}^{H} \bbx$. We can model this operation as a direct problem, where the input is given by $\bbx$, the linear transform is $\bbH = \bbV_{k}^{H}$ and the output is $\bby = \tbx_{k}$. We can thus proceed to reduce the complexity of this operation by designing a sampling pattern $\bbC$ and a matrix sketch $\bbH_{s}$ that compute an approximate output operating only on a subset of $p \ll n$ samples of $\bbx$ [cf. \eqref{eqn_direct_sketching}]. This way, the computational complexity is reduced by a factor of $p/n$ when compared to computing the GFT using $\bbV_{k}^{H}$ directly on the entire graph signal $\bbx$.

We consider a substantially larger problem with an Erd\H{o}s-R\'{e}nyi (ER) graph $\ccalG_{\textrm{ER}}$ of $n=10,000$ nodes and where edges connecting pairs of nodes are drawn independently with $p_{\textrm{ER}}=0.1$. The signal under study is bandlimited with $k=10$ frequency coefficients, generated in the same way as in Section~\ref{subsec_gft_sims}. To solve the problem \eqref{eqn_prob1_forward} we consider the NAH in a3), the NBH in a4) and the greedy approach in a5). We also consider the case in which the linear transform is sampled directly \eqref{eqn_operator_direct_sketching}, solving \eqref{eqn_operator_direct_sketching_optimal_matrix} by the greedy approach. Comparisons are carried out against all the methods in b).

%
\begin{figure}[t]
    \captionsetup[subfigure]{justification=centering}
    \centering
    \begin{subfigure}{0.49\textwidth}
        \centering
        \includegraphics[width=0.99\textwidth]{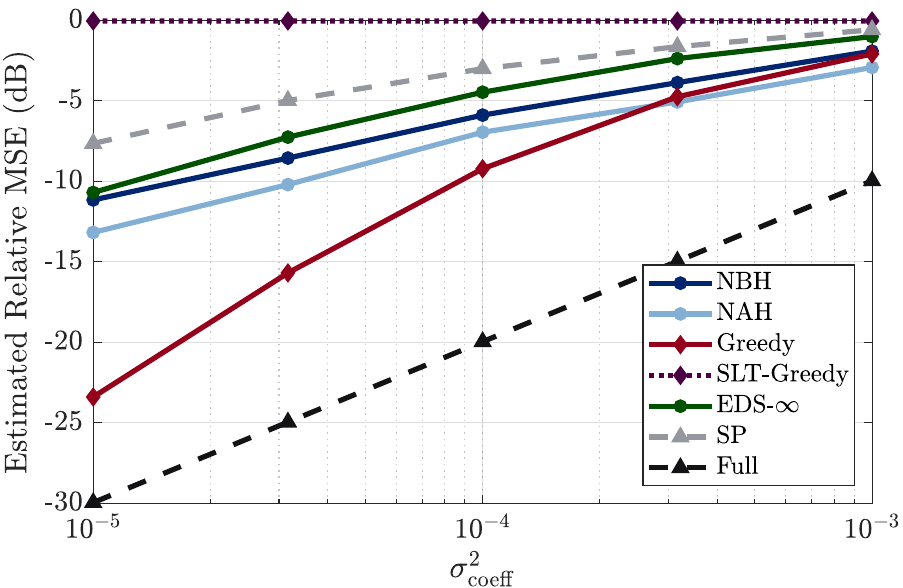}
        \caption{Function of the noise variance}
        \label{ER-noise}
    \end{subfigure}
    \hfill
    \begin{subfigure}{0.49\textwidth}
        \centering
        \includegraphics[width=0.99\columnwidth]{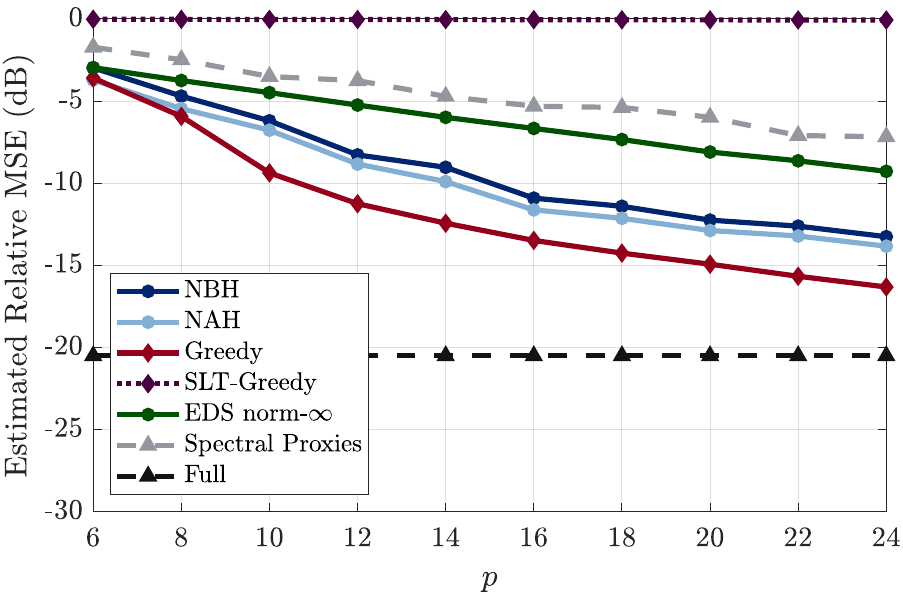}
        \caption{Function of the sample size}
        \label{ER-p}
    \end{subfigure}
    \caption{Approximating the GFT as a direct sketching problem for a large network. Relative estimated MSE $\|\hby-\tbx_{k}\|^{2}/\|\tbx_{k}\|^{2}$ for the problem of estimating the $k=10$ frequency components of a bandlimited graph signal from noisy observations, supported on an ER graph with $p=0.1$ and size $n=10,000$. \subref{ER-noise} As a function of $\sigma^{2}_{\textrm{coeff}}$ for fixed $p=k=10$. \subref{ER-p} As a function of the $p$ for fixed $\sigma^{2}_{\textrm{coeff}}=10^{-4}$.}
    \label{fig_direct_10k_10}
\end{figure}

In Figure~\ref{fig_direct_10k_10} we show the relative MSE for each method, in two different simulations. First, in Figure~\ref{ER-noise} the simulation was carried out as a function of noise $\sigma^{2}_{\textrm{coeff}}$ varying from $10^{-5}$ to $10^{-3}$, for a fixed number of samples $p=k=10$. We observe that the greedy approach in a5) performs best. The NAH in a3) and the NBH in a4) are the next best performers, achieving a comparable MSE. The EDS-$\infty$ in b1) yields the best results among the competing methods. We see that for the low-noise case ($\sigma^{2}_{\textrm{coeff}} = 10^{-5}$) we can obtain a performance that is $7\text{dB}$ worse than the baseline, but using only $p=10$ nodes out of $n=10,000$, thereby \emph{reducing the online computational cost of computing the GFT coefficients by a factor of $1,000$}.

For the second simulation, whose results are shown in Figure~\ref{ER-p}, we fixed the noise at $\sigma^{2}_{\textrm{coeff}}=10^{-4}$ and varied the number of samples from $p=6$ to $p=24$. Again, the greedy approach in a5) outperforms all the other methods, and the NAH in a3) and the NBH in a4) as the next best performers. The EDS-$\infty$ in b1) performs better than the SP method in b2). We note that, for $p<k$, the EDS-$\infty$ method has a performance very close to the NBH in a4), but as $p$ grows larger, the gap between them widens, improving the relative performance of the NBH. When selecting $p=24$ nodes, the greedy approach achieves an MSE that is $4\text{dB}$ worse than the baseline, but incurring in only $0.24\%$ of the online computational cost.

With respect to the off-line cost, the EDS-$\infty$ method incurs in $O(n^{3}+n^{2}\log n + n \log n)$ which, in this setting is $O(10^{12})$, while the greedy cost is $O(10^{11})$; yet, the MSE of the greedy approach is over an order of magnitude better than that of EDS-$\infty$. Likewise, the NAH and the NBH yield a lower, but comparable performance, with an off-line cost of $O(10^{9})$ and $O(10^{6})$, respectively.

\FloatBarrier

%
\subsection{Sensor selection for distributed parameter estimation} \label{subsec_sensor}

Here we address the problem of sensor selection for communication-efficient distributed parameter estimation~\cite{Geert}. The model under study considers the measurements $\bbx \in \reals^{n}$ of each of the $n$ sensors to be given as a linear transform of some unknown parameter $\bby \in \reals^{m}$, $\bbx = \bbH^{T} \bby$, where $\bbH^{T} \in \reals^{n \times m}$ is the observation matrix; refer to \cite[Section II]{Geert} for details.

We consider $n=96$ sensors, $m=12$ unknown parameters, and that the bandwidth of the sensor measurements is $k=10$. Following \cite[Section VI-A]{Geert}, matrix $\bbH^{T}$ is random where each element is drawn independently from a zero-mean Gaussian distribution with variance $1/\sqrt{n}$. The underlying graph support $\ccalG_{\textrm{U}}$ is built as follows. Each sensor is positioned at random, according to a uniform distribution, in the region $[0,1]^{2}$ of the plane. With $d_{i,j}$ denoting the Euclidean distance between sensors $i$ and $j$, their corresponding link weight is computed as $w_{ij} = \alpha e^{-\beta d^{2}_{i,j}}$, where $\alpha$ and $\beta$ are constants selected such that the minimum and maximum weights are $0.01$ and $1$. The network is further sparsified by keeping only the edges in the $4$-nearest neighbor graph. Finally, the resulting adjacency matrix is used as the graph shift operator $\bbS$.

%
\begin{figure}[t]
    \captionsetup[subfigure]{justification=centering}
    \centering
    \begin{subfigure}{0.49\textwidth}
        \centering
        \includegraphics[width=0.99\textwidth]{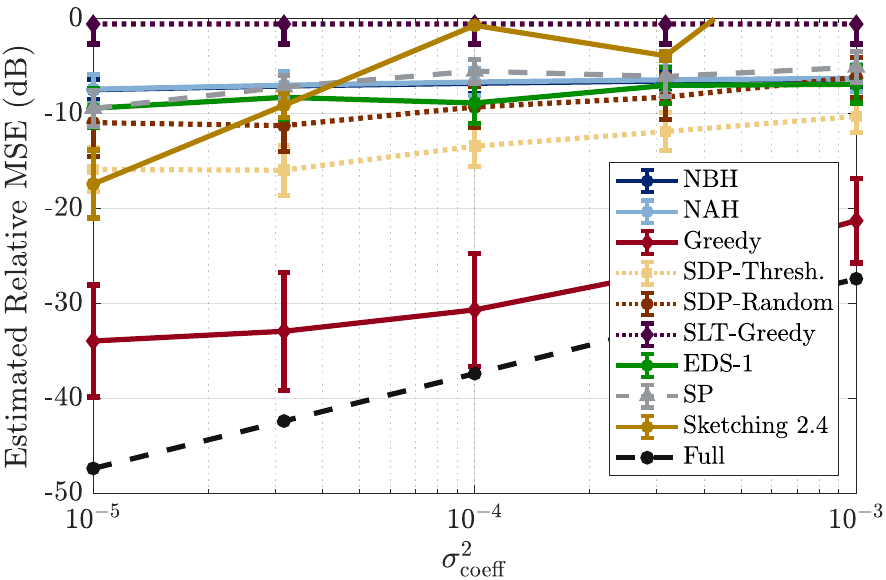}
        \caption{Function of the noise variance}
        \label{sensor-noise}
    \end{subfigure}
    \hfill
    \begin{subfigure}{0.49\textwidth}
        \centering
        \includegraphics[width=0.99\textwidth]{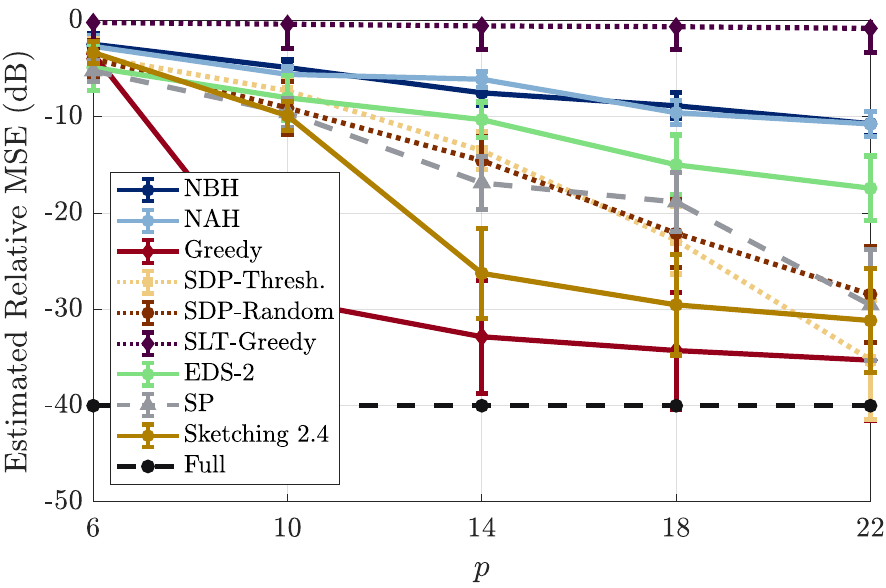}
        \caption{Function of the sample size}
        \label{sensor-p}
    \end{subfigure}
    \caption{Sensor selection for parameter estimation. Sensor are distributed uniformly over the $[0,1]^{2}$ region of the plane. Sensor graph is built using a Gaussian kernel over the Euclidean distance between sensors, and keeping only $4$ nearest neighbors. Relative estimated MSE as: \subref{sensor-noise} A function of $\sigma^{2}_{\textrm{coeff}}$ and \subref{sensor-p} A function of $p$. Legends: SDP-Random (scheme in a1); SDP-Thresh. (scheme in a2); NAH (scheme in a3); NBH (scheme in a4); Greedy (scheme in a5); SLT, EDS-1 and EDS-2 (schemes in b1); SP (scheme in b2); Sketching 2.6, Sketching 2.11 (schemes in c); optimal solution using the full signal.}
    \label{fig_sensor}
\end{figure}

For the simulations in this setting we generate a collection $\{\bbx_{t}+\bbw_{t}\}_{t=1}^{100}$ of sensor measurements. For each one of these measurements, $100$ noise realizations are drawn to estimate the relative MSE defined as $\mbE[\|\bby-\hby\|^{2}]/\mbE[\|\bby\|^{2}]$. Recall that $\hby = \bbH_{s}^{\ast}\bbC^{\ast}(\bbx+\bbw)$ [cf. \eqref{eqn_inverse_sketching}], with $\bbH_{s}^{\ast}$ given in Proposition~\ref{prop_noisy_forward} and $\bbC^{\ast}$ designed according to the methods under study. We run the simulations for $5$ different sensor networks and average the results, which are shown in Figure~\ref{fig_sensor}.

For the first simulation, the number of samples is fixed as $p=k=10$ and the noise coefficient $\sigma^{2}_{\textrm{coeff}}$ varies from $10^{-5}$ to $10^{-3}$. The estimated relative MSE is shown in Figure~\ref{sensor-noise}. We observe that the greedy approach in a5) performs considerably better than any other method. The solution provided by SDP-Thresh. in a2) exhibits the second best performance. With respect to the  methods under comparison, we note that EDS-1 in b1) is the best performer and outperforms NAH in a3) and the NBH in a4), but is still worse than SDP-Random in a1), although comparable. We also observe that while traditional Sketching 2.4 in c) yields good results for low-noise scenarios, its performance quickly degrades as the noise increases. We conclude that using the greedy approach it is possible to estimate the parameter $\bby$ with a $4\text{dB}$ loss, with respect to the optimal, baseline solution, but taking measurements from only $10$ out of the $96$ deployed sensors.

In the second simulation, we fixed the noise given by $\sigma^{2}_{\textrm{coeff}} = 10^{-4}$ and varied the number of samples from $p=6$ to $p=22$. The estimated relative MSE is depicted in Figure~\ref{sensor-p}. We observe that the greedy approach in a5) outperforms all other methods. We also note that, in this case, the Sketching 2.4 algorithm in c) has a very good performance. It is also worth pointing out that the SP method in b2) outperforms the EDS scheme in b1), and that the performance of the SDP relaxations in a1) and a2) improves considerably as $p$ increases.

\FloatBarrier

%
\subsection{MNIST handwritten digits classification} \label{subsec_digits}

Images are another example of signals that (approximately) belong to a lower-dimensional subspace. In fact, principal component analysis (PCA) shows that only a few coefficients are enough to describe an image \cite{pca, shahid17-pcagraph}. More precisely, if we vectorize an image, compute its covariance matrix, and project it onto the eigenvectors of this matrix, then the resulting vector (the PCA coefficients) would have most of its components almost zero. This shows that natural images are approximately bandlimited in the subspace spanned by the eigenvectors of the covariance matrix, and thus are suitable for sampling.

We focus on the problem of classifying images of handwritten digits from the MNIST database \cite{mnist}. To do so, we use a linear support vector machine (SVM) classifier \cite{jain89}, trained to operate on a few of the PCA coefficients of each image. We can model this task as a direct problem where the linear transform to apply to each (vectorized) image is the cascade of the projection onto the covariance matrix eigenvectors (the PCA transform), followed by the linear SVM classifier.

To be more formal, let $\bbx \in \reals^{n}$ be the vectorized image, where $n = 28 \times 28 = 784$ is the total number of pixels. Each element of this vector represents the value of a pixel. Denote by $\bbR_{x} = \bbV \bbLambda \bbV^{T}$ the covariance matrix of $\bbx$. Then, the PCA coefficients are computed as $\bbx^{\textrm{PCA}} = \bbV^{T} \bbx$ [cf. \eqref{eqn_projection}]. Typically, there are only a few non-negligible PCA coefficients which we assume to be the first $k \ll n$ elements. These elements can be directly computed by $\bbx_{k}^{\textrm{PCA}} = \bbV_{k}^{T} \bbx$ [cf. \eqref{eqn_recovery}]. Then, these $k$ PCA coefficients are fed into a linear SVM classifier, $\bbA_{\textrm{SVM}} \in \reals^{m \times k}$ where $m$ is the total number of digits to be classified (the total number of classes). Lastly, $\bby = \bbA_{\textrm{SVM}} \bbx_{k}^{\textrm{PCA}} = \bbA_{\textrm{SVM}} \bbV_{k}^{T} \bbx$ is used to determine the class (typically, by assigning the image to the class corresponding to the maximum element in $\bby$). This task can be cast as a direct sketching problem \eqref{eqn_direct_sketching}, by assigning the linear transform to be $\bbH = \bbA_{\textrm{SVM}} \bbV_{k}^{T} \in \reals^{m \times n}$, with $\bby \in \reals^{m}$ being the output and the vectorized image $\bbx \in \reals^{n}$ being the input.

%
\begin{figure}[t]
    \captionsetup[subfigure]{justification=centering}
    \centering
    \begin{subfigure}{0.24\columnwidth}
        \centering
        \includegraphics[width=\digitsize\textwidth]{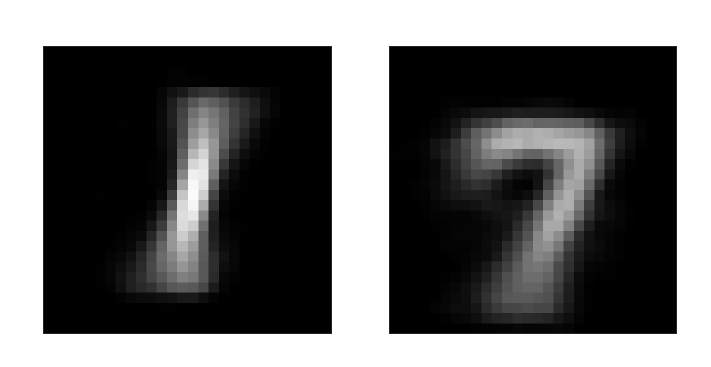}
        \caption{Full\\image\\($1.00\%$)}
        \label{digits-full}
    \end{subfigure}
    \hfill
    \begin{subfigure}{0.24\columnwidth}
        \centering
        \includegraphics[width=\digitsize\textwidth]{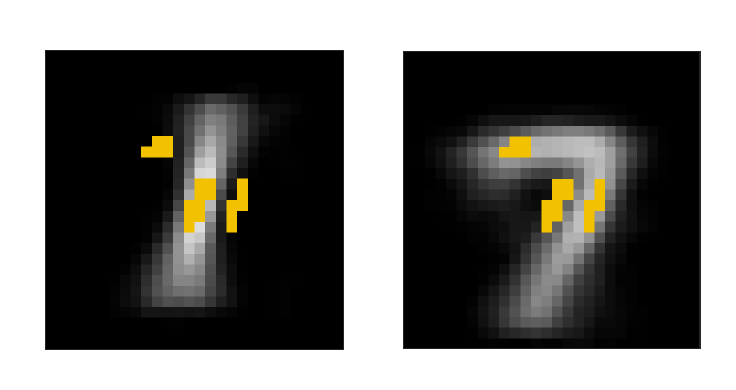}
        \caption{Noise-Blind Heuristic\\($3.27\%$)}
        \label{digits-nbh}
    \end{subfigure}
    \hfill
    \begin{subfigure}{0.24\textwidth}
        \centering
        \includegraphics[width=\digitsize\textwidth]{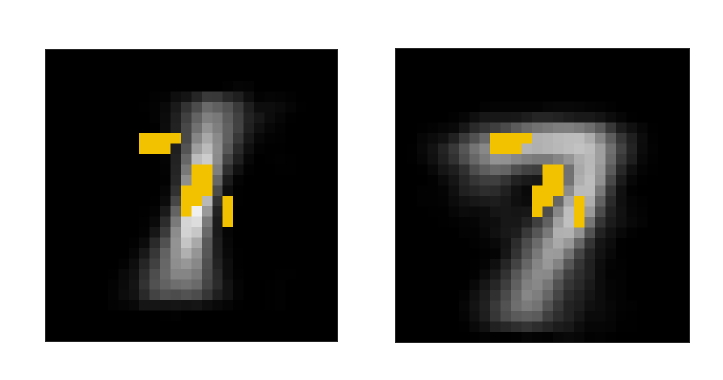}
        \caption{Noise-Aware Heuristic\\($2.81\%$)}
        \label{digits-nah}
    \end{subfigure}
    \hfill
    \begin{subfigure}{0.24\textwidth}
        \centering
        \includegraphics[width=\digitsize\textwidth]{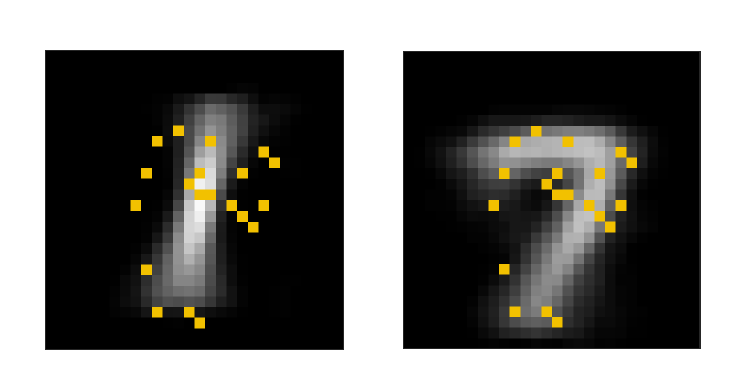}
        \caption{Greedy\\Approach\\($1.39\%$)}
        \label{digits-greedy}
    \end{subfigure}

    \begin{subfigure}{0.24\textwidth}
        \centering
        \includegraphics[width=\digitsize\textwidth]{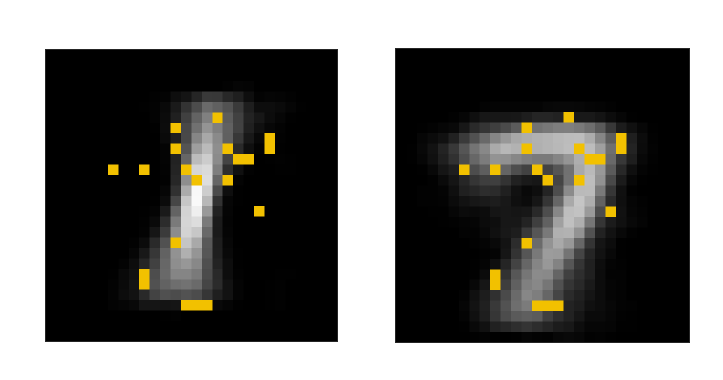}
        \caption{SDP Relaxation -- Thresholding\\($2.73\%$)}
        \label{digits-crt}
    \end{subfigure}
    \hfill
    \begin{subfigure}{0.24\textwidth}
        \centering
        \includegraphics[width=\digitsize\textwidth]{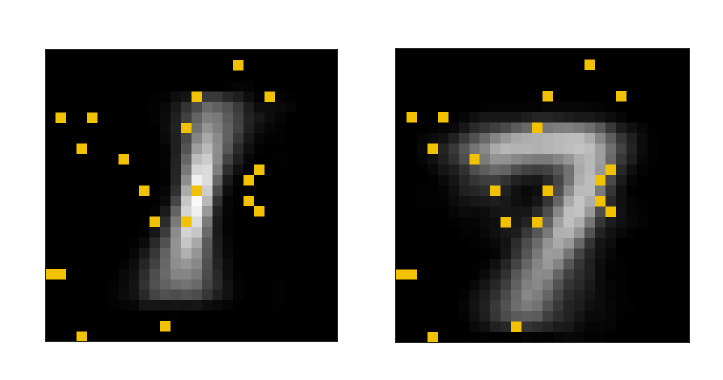}
        \caption{SDP Relaxation -- Random\\($3.55\%$)}
        \label{digits-crr}
    \end{subfigure}
    \hfill
    \begin{subfigure}{0.24\columnwidth}
        \centering
        \includegraphics[width=\digitsize\textwidth]{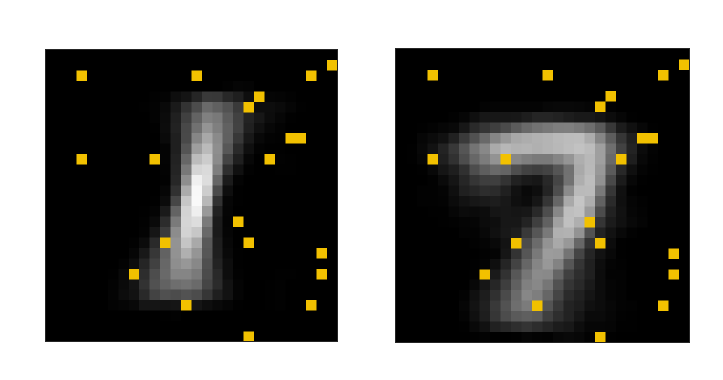}
        \caption{Sampling Linear Transform\\($5.65\%$)}
        \label{digits-scg}
    \end{subfigure}
    \hfill
    \begin{subfigure}{0.24\columnwidth}
        \centering
        \includegraphics[width=\digitsize\textwidth]{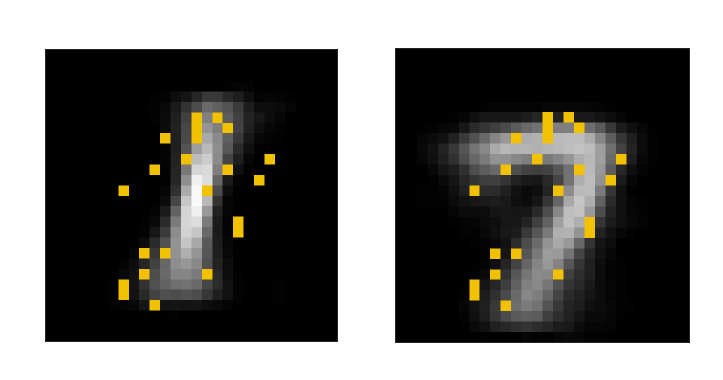}
        \caption{EDS\\norm-$1$\\($3.70\%$)}
        \label{digits-eds}
    \end{subfigure}
    
    \caption{Selected pixels to use for classification of the digits according to each strategy. The percentage error (ratio of errors to total number of images) is also shown. \subref{digits-full} Images showcasing the average of all images in the test set labeled as a \texttt{1} (left) and as a \texttt{7} (right). Using all pixels to compute $k=20$ PCA coefficients and feeding them to a linear SVM classifier yields $1.00\%$ error.}
    \label{fig_digits} 
\end{figure}

In this experiment, we consider the classification of $c$ different digits. The MNIST database consists of a training set of $60,000$ images and a test set of $10,000$. We select, uniformly at random, a subset $\ccalT$ of training images and a subset $\ccalS$ of test images, containing only images of the $c$ digits to be classified. We use the $|\ccalT|$ images in the training set to estimate the covariance matrix $\bbR_{x}$ and to train the SVM classifier \cite{duda01}. Then, we run on the test set $\ccalS$ the classification using the full image, as well as the results of the sketching-as-sampling method implemented by the five different heuristics in a), and compute the classification error $e = |\# \textrm{misclassified images}|/|\ccalS|$ as a measure of performance. The baseline method, in this case, stems from computing the classification error obtained when operating on the entire image, without using any sampling or sketching. For all simulations, we add noise to the collection of images to be classified $\{\bbx_{t}+\bbw_{t}\}_{t=1}^{|\ccalS|}$, where $\bbw_{t}$ is zero-mean Gaussian noise with variance $\bbR_{w} = \sigma_{w}^{2} \bbI_{n}$, with $\sigma_{w}^{2} = \sigma_{\textrm{coeff}}^{2} \mbE[\|\bbx\|^{2}]$ (where the expected energy is estimated from the images in the training set $\ccalT$). We assess performance as function of $\sigma^{2}_{\textrm{coeff}}$ and $p$.

%
\begin{figure}[t]
    \captionsetup[subfigure]{justification=centering}
    \centering
    
    \begin{subfigure}{0.49\textwidth}
        \centering
        \includegraphics[width=0.99\textwidth]{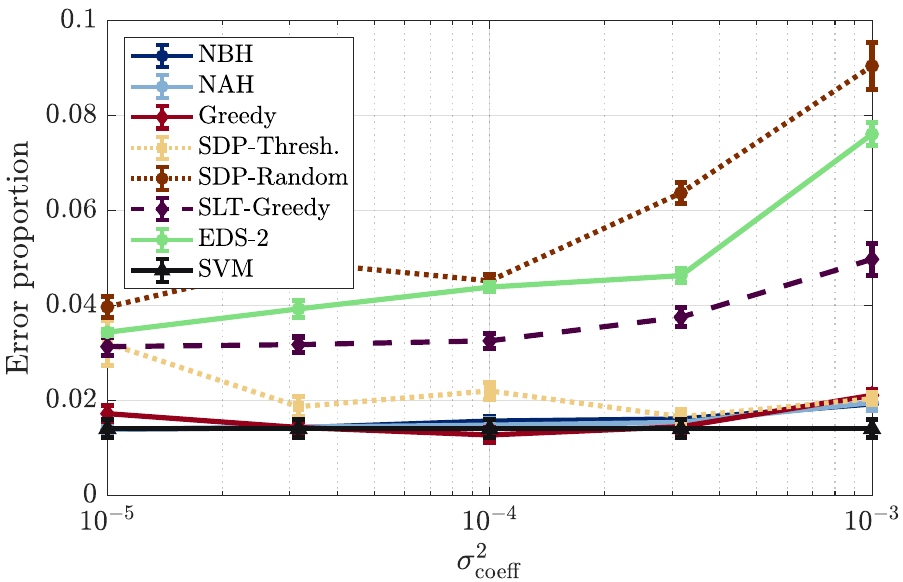}
        \caption{$\{\texttt{1},\texttt{7}\}$: noise}
        \label{two-digits-noise}
    \end{subfigure}
    \hfill
    \begin{subfigure}{0.49\textwidth}
        \centering
        \includegraphics[width=0.99\textwidth]{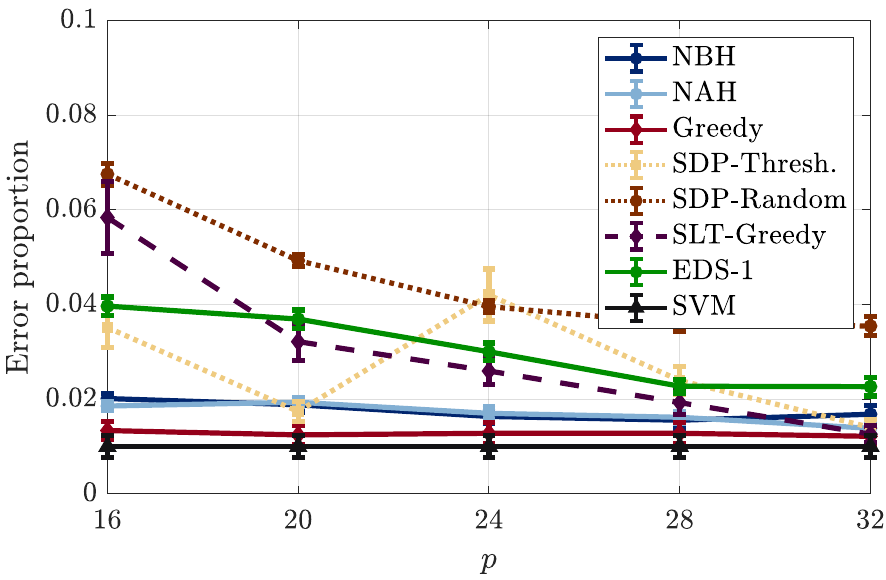}
        \caption{$\{\texttt{1},\texttt{7}\}$: number of samples}
        \label{two-digits-p}
    \end{subfigure}
    \caption{MNIST Digit Classification for digits $\texttt{1}$ and $\texttt{7}$. Error proportion as \subref{two-digits-noise} a function of noise $\sigma^{2}_{\textrm{coeff}}$ and as \subref{two-digits-p} a function of the number of samples $p$. We note that in both cases the greedy approach in a5) works best. The worst case difference between the greedy approach and the SVM classifier using the entire image is $0.33\%$, which happens when attempting classification with just $16$ pixels out of $784$.}
    \label{fig_mnist_two-digits}
\end{figure}

We start by classifying digits $\{\texttt{1},\texttt{7}\}$, that is $c=2$. We do so for fixed $p=k=20$ and $\sigma^{2}_{\textrm{coeff}}=10^{-4}$. The training set has $|\ccalT|=10,000$ images ($5,000$ of each digit) and the test set contains $|\ccalS|=200$ images ($100$ of each). Figure~\ref{fig_digits} illustrates the averaged images of both digits (averaged across all images of each given class in the test set $\ccalS$) as well as the selected pixels following each different selection technique. We note that, when using the full image, the percentage error obtained is $1\%$, which means that $2$ out $|\ccalS|=200$ images were misclassified. The greedy approach (Figure~\ref{digits-greedy}) has a percentage error of $1.5\%$ which entails $3$ misclassified images, only one more than the full image SVM classifier, but using only $p=20$ pixels instead of the $n=784$ pixels of the image. Remarkably, even after reducing the online computational cost by a factor of $39.2$, we only incur a marginal performance degradation (a single additional misclassified image). Moreover, solving the direct sketching-as-sampling problem with any of the proposed heuristics in a) outperforms the selection matrix obtained by EDS-$1$. When considering the computationally simpler problem of using the same sampling matrix for the signal and the linear transform [cf. \eqref{eqn_operator_direct_sketching}], the error incurred is of $5.65\%$. Finally, we note that the sketching-as-sampling techniques tend to select pixels for classification that are different in each image (pixels that are black in the image of one digit and white in the image of the other digit, and vice versa), i.e., the most discriminative pixels.

%
\begin{figure}[t]
    \captionsetup[subfigure]{justification=centering}
    \centering
    \begin{subfigure}{0.49\textwidth}
        \centering
        \includegraphics[width=0.99\textwidth]{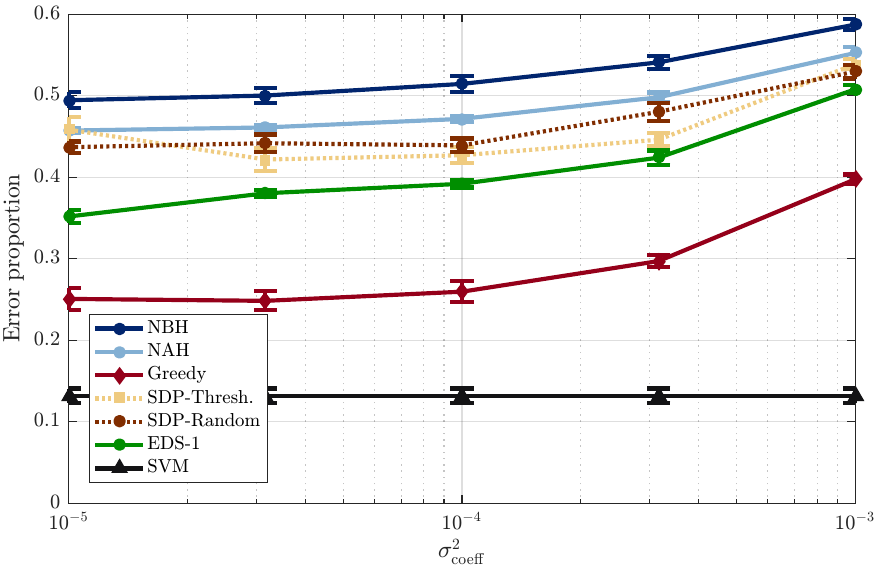}
        \caption{$\{\texttt{0},\ldots,\texttt{9}\}$: noise}
        \label{ten-digits-noise} 
    \end{subfigure}
    \hfill
    \begin{subfigure}{0.49\textwidth}
        \centering
        \includegraphics[width=0.99\textwidth]{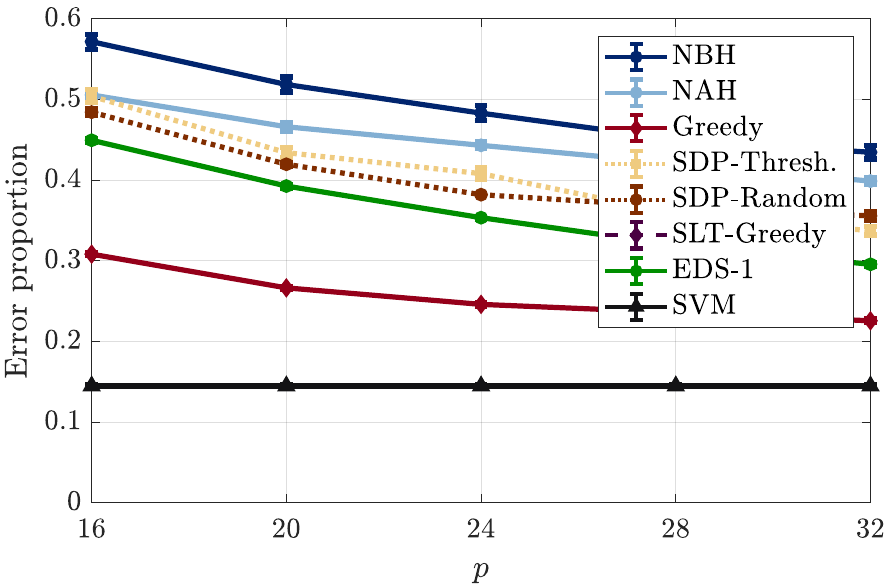}
        \caption{$\{\texttt{0},\ldots,\texttt{9}\}$: number of samples}
        \label{ten-digits-p} 
    \end{subfigure}
    
    \caption{MNIST Digit Classification for all ten digits. Error proportion as \subref{ten-digits-noise} a function of noise $\sigma^{2}_{\textrm{coeff}}$ and as \subref{ten-digits-p} a function of the number of samples $p$. We note that in both cases the greedy approach in a5) works best. The worst case difference between the greedy approach and the SVM classifier using the entire image is $16.36\%$.}
    \label{fig_mnist_ten-digits}
\end{figure}

For the first parametric simulation, we consider the same two digits $\{\texttt{1},\texttt{7}\}$ under the same setting as before, but, in one case, for fixed $p=k=20$ and varying noise $\sigma^{2}_{\textrm{coeff}}$ (Figure~\ref{two-digits-noise}) and for fixed $\sigma^{2}_{\textrm{coeff}} = 10^{-4}$ and varying $p$ (Figure~\ref{two-digits-p}). We carried out these simulations for $5$ different random dataset train/test splits. The greedy approach in a5) outperforms all other solutions in a), and performs comparably to the SVM classifier using the entire image. The NAH and NBH also yield satisfactory performance. In the most favorable situation, the greedy approach yields the same performance as the SVM classifier on the full image, but using only $20$ pixels, the worst case difference is of $0.33\%$ ($1$ image) when using only $p=16$ pixels ($49\%$ reduction in computational cost).

For the second parametric simulation, we consider $c=10$ digits: $\{\texttt{0},\texttt{1},\ldots,\texttt{9}\}$ (Figure~\ref{fig_mnist_ten-digits}). We observe that the greedy approach in a5) is always the best performer, although the relative performance with respect to the SVM classifier on the entire image worsens. We also observe that the NAH and NBH are more sensitive to noise, being outperformed by the EDS-1 in b1) and the SDP relaxations. These three simulations showcase the tradeoff between faster computations and performance.

%
\subsection{Authorship attribution} \label{subsec_author}

As a last example of the sketching-as-sampling methods, we address the problem of authorship attribution where we want to determine whether a text was written by a given author or not \cite{Segarra15-WANs}. To this end, we collect a set of texts that we know have been written by the given author (the training set), and build a word adjacency network (WAN) of function words for each of these texts. Function words, as defined in linguistics, carry little lexical meaning or have ambiguous meaning and express grammatical relationships among other words within a sentence, and as such, cannot be attributed to a specific text due to its semantic content. It has been found that the order of appearance of these function words, together with their frequency, determine a stylometric signature of the author. WANs capture, precisely, this fact. More specifically, they determine a relationship between words using a mutual information measure based on the order of appearance of these words (how often two function words appear together and how many other words are usually in between them). For more details on function words and WAN computation, please refer to \cite{Segarra15-WANs}.

\begin{figure}[t]
    \captionsetup[subfigure]{justification=centering}
    \centering
    \begin{subfigure}{0.49\textwidth}
        \centering
        \includegraphics[width=0.99\textwidth]{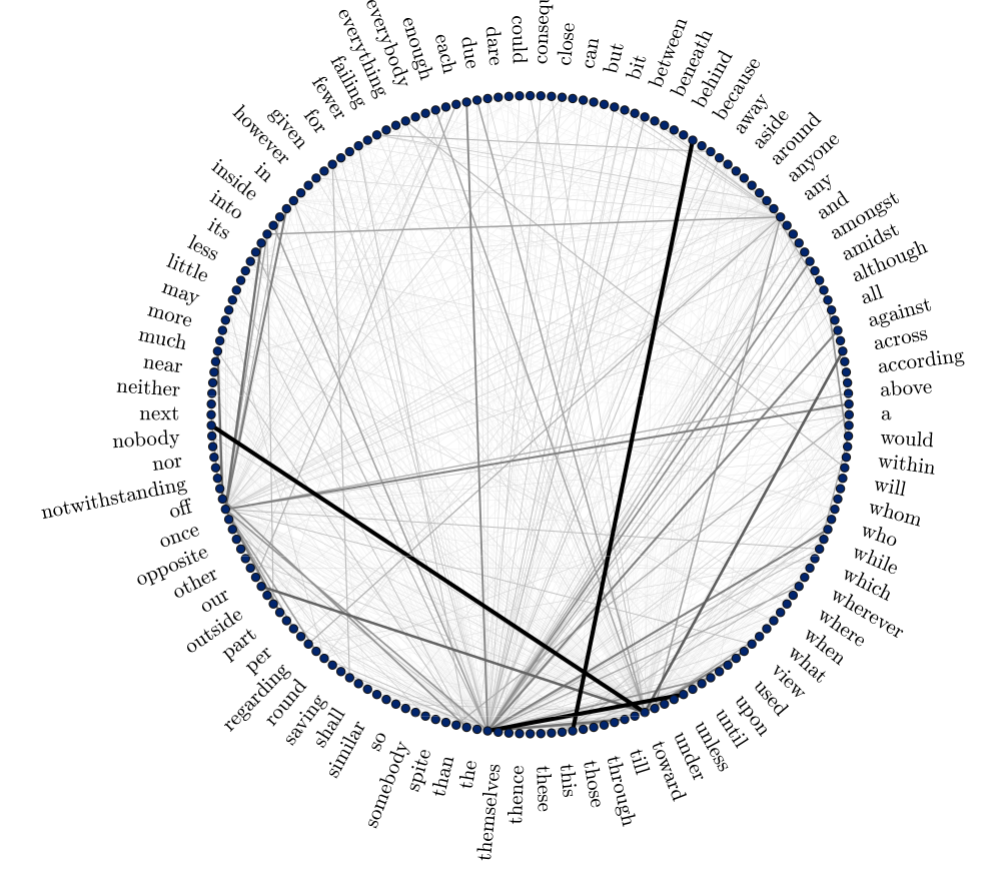}
        \caption{Word adjacency network}
        \label{wan-all} 
    \end{subfigure}
    \hfill
    \begin{subfigure}{0.49\textwidth}
        \centering
        \includegraphics[width=0.99\textwidth]{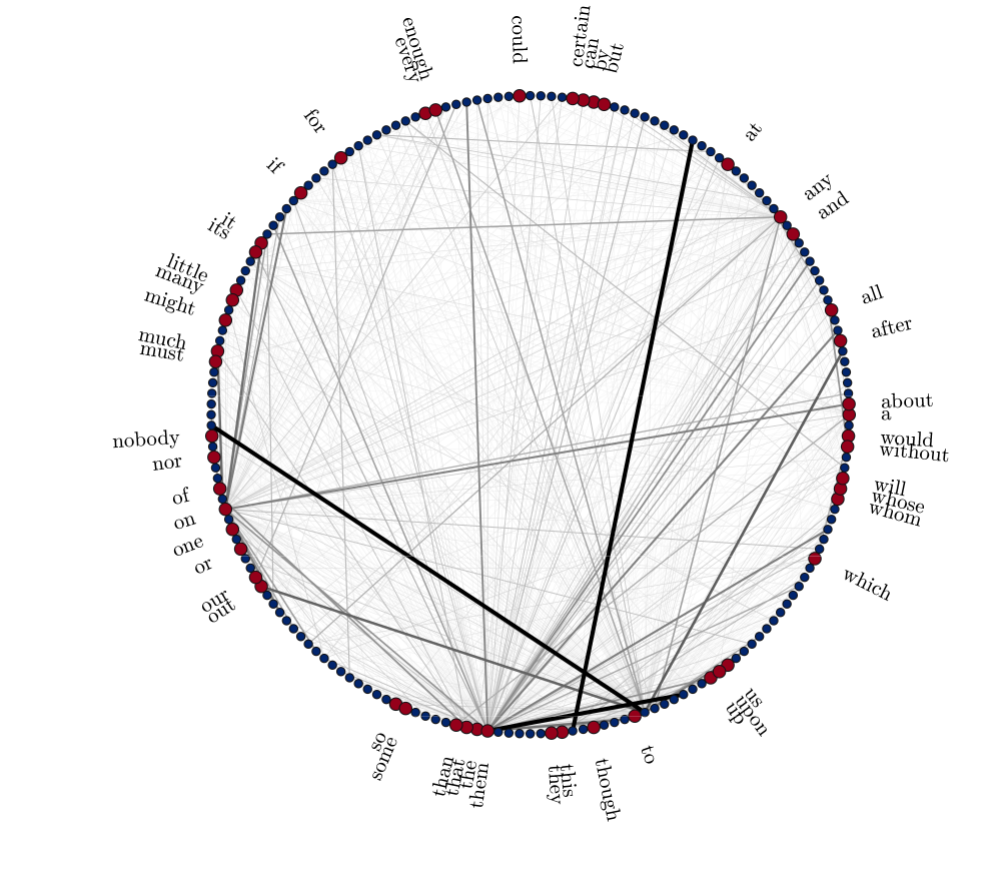}
        \caption{Selected words}
        \label{wan-greedy} 
    \end{subfigure}
    
    \caption{Word adjacency networks (WANs). \subref{wan-all} Example of a WAN built from the training set of texts written by Jane Austen; to avoid clutter only one every other word are shown as node labels. \subref{wan-greedy} Highlighted in red are the words selected by the greedy approach in a5).}
    \label{fig_wan}
\end{figure}

In what follows, we consider the corpus of novels written by Jane Austen. Each novel is split in fragments of around $1,000$ words, leading to $771$ texts. Of these, we take $617$ at random to be part of the training set, and $154$ to be part of the test set. For each of the $617$ texts in the training set we build a WAN considering $211$ function words, as detailed in \cite{Segarra15-WANs}. Then we combine these WANs to build a single graph, undirected, normalized and connected (usually leaving around $190$ function words for each random partition, where some words were discarded to make the graph connected). An illustration of one realization of a resulting graph can be found in Figure~\ref{fig_wan}. The adjacency matrix of the resulting graph is adopted as the shift operator $\bbS$.

%

Now that we have built the graph representing the stylometric signature of Jane Austen, we proceed to obtain the corresponding graph signals. We obtain the word frequency count of the function words on each of the $771$ texts, respecting the split $617$ for training and $154$ for test set, and since each function word represents a node in the graph, the word frequency count can be modeled as a graph signal. The objective is to exploit the relationship between the graph signal (word frequency count) and the underlying graph support (WAN) to determine whether a given text was authored by Jane Austen or not. To do so, we use a linear SVM classifier (in a similar fashion as in the MNIST example covered in Section~\ref{subsec_digits}). The SVM classifier is trained by augmenting the training set with another $617$ graph signals (word frequency count) belonging to texts written by other contemporary authors, including Louisa May Alcott, Emily Bront\"{e}, Charles Dickens, Mark Twain, Edith Wharton, among others. The $617$ graph signals corresponding to texts by Jane Austen are assigned a label $\texttt{1}$ and the remaining $617$ samples are assigned a label $\texttt{0}$. This labeled training set of $1,234$ samples is used to carry out supervised training of a linear SVM. The trained linear SVM serves as a linear transform on the incoming graph signal, so that it becomes $\bbH$ in the direct model problem. We then apply the sketching-as-sampling methods discussed in this work to each of the texts in the test set (which has been augmented to include $154$ texts of other contemporary authors, totaling $308$ samples). To assess performance,  we evaluate the error rate of determining whether the texts in the test set were authored by Jane Austen or not, that is achieved by the \emph{sketched} linear classifiers operating on only a subset of function words (selected nodes).

\begin{figure}[t]
    \captionsetup[subfigure]{justification=centering}
    \centering
    \begin{subfigure}{0.49\textwidth}
        \centering
        \includegraphics[width=0.99\textwidth]{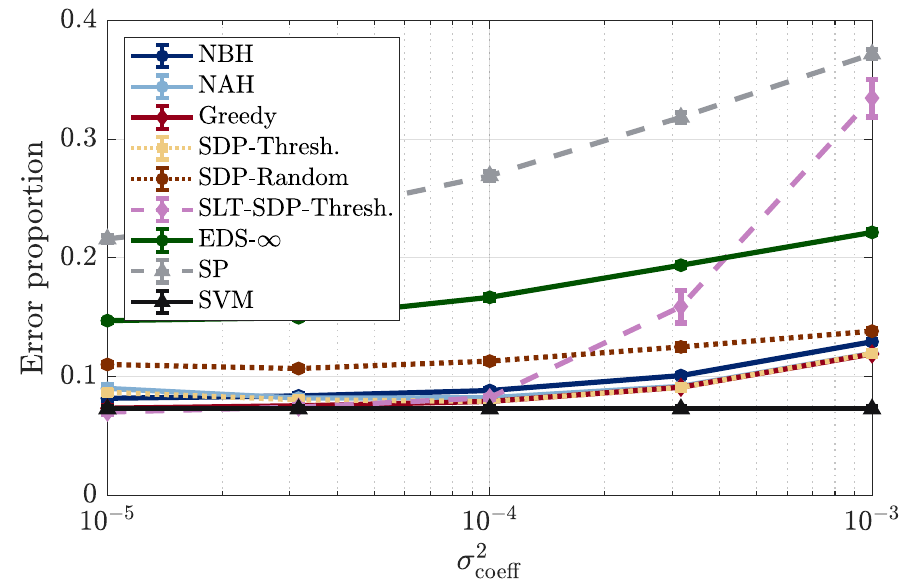}
        \caption{noise}
        \label{author-noise} 
    \end{subfigure}
    \hfill
    \begin{subfigure}{0.49\textwidth}
        \centering
        \includegraphics[width=0.99\textwidth]{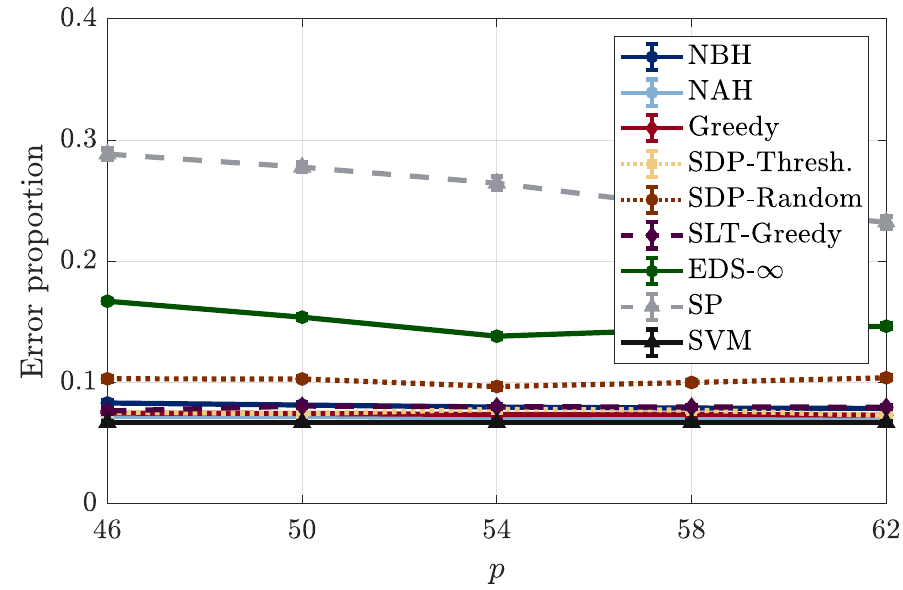}
        \caption{number of samples}
        \label{author-p} 
    \end{subfigure}
    
    \caption{Authorship attribution for texts written by Jane Austen. Error proportion as \subref{author-noise} a function of noise $\sigma^{2}_{\textrm{coeff}}$ and as \subref{author-p} a function of the number of samples $p$ (number of words selected). We note that in both cases the greedy approach in a5), the noise-aware heuristic (NAH) in and the SDP relaxation with thresholding (SDP-Thresh.) in offer similar, best performance. The baseline error for an SVM operating on the full text is $7.27\%$ in \subref{author-noise} and $6.69\%$ in \subref{author-p}.}
    \label{fig_author}
\end{figure}

For the experiments, we thus considered sequences $\{\bbx_{t}\}_{t=1}^{308}$ of the $308$ test samples, where each $\bbx_{t}$ is the graph signal representing the word frequency count of each text $t$ in the test set. After observing the graph frequency response of these graph signals over the graph (given by the WAN), we note that signals are approximately bandlimited with $k=50$ components. We repeat the experiment for $5$ different realizations of the random train/test set split of the corpus. To estimate $\bbR_{x}$ we use the sample covariance of the graph signals in the training set. The classification error is computed as the proportion of mislabeled texts out of the $308$ test samples, and is averaged across the random realizations of the dataset split. We run simulations for different noise levels, computed as $\sigma_{w}^{2} = \sigma_{\textrm{coeff}}^{2} \cdot \mbE[ \| \bbx \|^{2}]$ for a fixed number of $p=k=50$ selected nodes (function words), and also for different number of selected nodes $p$ for a fixed noise coefficient $\sigma_{\textrm{coeff}}^{2}=10^{-4}$. Results are shown in Figure~\ref{fig_author}. Additionally, Figure~\ref{fig_wan} illustrates an example of the selected words for one of the realizations.

%

In Figure~\ref{author-noise} we show the error rate as a function of noise for all the methods considered. First and foremost, we observe a baseline error rate of $7.27\%$ corresponding to the SVM acting on all function words. As expected, observe that the performance of all the methods degrades as more noise is considered (classification error increases). In particular, the greedy approach attains the lowest error rate, with a performance matched by both the SDP-relaxation with thresholding in a2) and the NAH in a3). It is interesting to observe that the technique of directly sampling the linear transform and the signal (method a6) performs as well as the greedy approach in the low-noise scenario, but then degrades rapidly. Finally, we note that selecting samples irrespective of the linear classifier exhibits a much higher error-rate than the sketching-as-sampling counterparts. This is the case for the EDS-$\infty$ shown in Figure~\ref{author-noise} (the best performer among all methods in b).

In the second experiment for fixed noise $\sigma_{\textrm{coeff}}^{2} = 10^{-4}$ and varying number of selected samples $p$, we show the resulting error rate in Figure~\ref{author-p}. The baseline for the SVM classifier using all the function words is a $6.69\%$ error rate. Next, we observe that the error rate is virtually unchanged as more function words are selected, suggesting that the performance of the methods in this example is quite robust. Among all the competing methods, we see that the best performing one is the NAH in a3), incurring in an error rate of $7\%$, but with comparable performance from the greedy approach in a5) as well as the SDP relaxation with thresholding in a2). The selection of words with complete disregard for the classification task (i.e. selecting words only for reconstruction) significantly degrades the performance, as evidenced by the error rate achieved by the EDS-$\infty$ (the best performer among all methods described in b).

\FloatBarrier

%
\section{Conclusions} \label{sec_conclusions}


In this work, we proposed a novel framework for sampling signals that belong to a lower-dimensional subspace. \blue{By viewing these signals as bandlimited graph signals, we leveraged notions of sampling drawn from the GSP framework}. The objective is to significantly reduce the cost of approximating the output of a linear transform of the given signal. This can be achieved by leveraging knowledge of both the signal statistics as well as the said linear operator.

Under the working principle that the linear transform has to be applied to a sequence of streaming signals, we jointly designed a sampling pattern together with a sketch of the linear operator so that this sketch can be affordably applied to a few samples of each incoming signal. The joint design problem was formulated as a two-stage optimization. We first found the expression for the optimal sketching matrix as a function of feasible sampling strategies, and then substituted this expression in the original MSE cost to solve for the optimal sampling scheme. Overall, the resulting \emph{sketched} output offers a good approximation to the output of the full linear transform, but at a much lower online computational cost.

We proposed five different heuristics to (approximately) solve the MSE minimization problems, which are intractable due to the (combinatorial) sample selection constraints. We then tested these heuristics, together with traditional sketching approaches and other sampling methods (that select samples for reconstruction) in four different scenarios, namely: computation of GFT coefficients, distributed estimation in a sensor network, handwritten digit classification and authorship attribution. All these GSP applications involve signals that belong to lower-dimensional subspaces. In general terms, we observed that the greedy heuristic works best, even though at a higher off-line computational cost. We also noted that the noise-blind and noise-aware heuristics offer reasonable performance at a very reduced cost, while the SDP relaxation performance is highly dependent on the application and does not scale gracefully for large problems. Any of these heuristics, however, outperforms both traditional sketching methods as well as sampling-for-reconstruction schemes in the recent GSP literature.

%
\section{Appendix} \label{sec_appendix}


For concreteness, here we give the proofs for the direct sketching problem. The arguments for the inverse problem are similar; see~\cite{FGAMGMAR_globalsip16} for details.

\subsection{Proof of Proposition \ref{prop_noisy_forward}} \label{proof_prop_noisy_forward}

%

The objective function in \eqref{eqn_direct_sketching_optimal_matrix} can be rewritten as

\begin{align} \label{eqn_obj_noisy_forward}
	   \mbE \left[ \| \bby - \hby \|_{2}^{2} \right]
	= {}& \mbE \left[ \| \bbH \bbx - \bbH_s \bbC (\bbx + \bbw) \|_{2}^{2} \right] \nonumber \\
	    = {}& \tr \left[ \bbH \bbR_{x}\bbH^{T} - 2 \bbH_s \bbC \bbR_{x} \bbH^{T} + \bbH_s \bbC (\bbR_{x}+\bbR_{w}) \bbC^{T} \bbH_s^{T} \right]  
\end{align}
since $\bbx$ and $\bbw$ are assumed independent. Optimizing with respect to $\bbH_s$ first, yields

\begin{equation}\label{eqn_sketching_matrix_forward}
	\bbH_s^{*}(\bbC) = \bbH \bbR_{x} \bbC^{T} \left( \bbC (\bbR_{x}+\bbR_{w}) \bbC^{T} \right)^{-1}
\end{equation}
establishing the first branch in \eqref{eqn_H2_noisy_forward}. Matrix $\bbC \in \ccalC$ is full rank since it selects $p$ \textit{distinct} samples, then $\bbC (\bbR_{x}+\bbR_{w}) \bbC^{T}$ has rank $p$ and thus it is invertible \cite{hornjohnson85}. Substituting \eqref{eqn_sketching_matrix_forward} into \eqref{eqn_obj_noisy_forward}, yields \eqref{eqn_prob1_forward}.\hfill$\Box$


\subsection{Proof of Proposition \ref{prop_noisy_forward_sdp}} \label{proof_prop_noisy_forward_sdp}
%

The inverse in the objective of \eqref{eqn_prob1_forward} can be written as~\cite{Geert,bova04}

\begin{align}
	   ( \bbC (\bbR_{x}+\bbR_{w}) \bbC^{T} )^{-1} 
	={}& \left( \alpha \bbI_{p} - \alpha \bbI_{p} + \bbC (\bbR_{x}+\bbR_{w}) \bbC^{T} \right)^{-1} \nonumber \\
	    ={}& \alpha^{-1} \bbI_{p}- \alpha^{-2} \bbC (\barbR_{\alpha}^{-1} + \alpha^{-1} \bbC^{T} \bbC )^{-1} \bbC^{T} \label{eqn_wmi_forward}
\end{align}
where $\alpha \ne 0$ is a rescaling parameter, and in obtaining the second equality we used the Woodbury Matrix Identity~\cite{golubvanloan85}. Note that $\alpha$ has to be such that $\barbR_{\alpha}=(\bbR_{x}+\bbR_{w}-\alpha \bbI_{n})$ is still invertible. Substituting \eqref{eqn_wmi_forward} into \eqref{eqn_prob1_forward} and recalling that $\bbC^{T}\bbC = \diag(\bbc)$, we have that

\begin{align}\label{eqn_prob1b_forward}
	\min_{\bbc \in \{0,1\}^{n},\barbC_{\alpha}}\
	& \tr \Big[ \bbH \bbR_{x} \bbH^{T} - \bbH \bbR_{x} \barbC_{\alpha} \bbR_{x} \bbH^{T} + \bbH \bbR_{x} \barbC_{\alpha}\left( \barbR_{\alpha}^{-1} + \barbC_{\alpha}\right)^{-1} \barbC_{\alpha} \bbR_{x} \bbH^{T} \Big]\nonumber \\
	\st\
	& \barbC_{\alpha}=\alpha^{-1}\diag(\bbc) \ , \ 
	\bbc^{T} \bbone_{n}=p .
\end{align}
Note that in \eqref{eqn_prob1b_forward} we optimize over a binary vector $\bbc \in \reals^{n}$ with exactly $p$ nonzero entries, instead of a binary matrix $\bbC \in \ccalC$. The $p$ nonzero elements in $\bbc$ indicate the samples to be taken. Problem \eqref{eqn_prob1b_forward} can be reformulated as

\begin{align}\label{eqn_prob1c_forward}
	\min_{\substack{\bbc \in \{0,1\}^{n},\\\bbY, \barbC_{\alpha}}}\
	& \tr \left[ \bbY \right] \\
	\st\
	& \bbH \bbR_{x} \bbH^{T} - \bbH \bbR_{x} \barbC_{\alpha} \bbR_{x} \bbH^{T}+ \bbH \bbR_{x} \barbC_{\alpha}\left( \barbR_{\alpha}^{-1} + \barbC_{\alpha}\right)^{-1} \barbC_{\alpha} \bbR_{x} \bbH^{T} \preceq \bbY \nonumber \\
	& \barbC_{\alpha}=\alpha^{-1}\diag(\bbc) \ , \
	\bbc^{T} \bbone_{n}=p
	\nonumber
\end{align}
where $\bbY \in \reals^{m \times m}$, $\bbY \succeq \mathbf{0}$ is an auxiliary optimization variable. Using the Schur-complement lemma for \textit{positive definite} matrices~\cite{bova04}, \eqref{eqn_prob1c_forward} can be written as \eqref{eqn_prob2_forward}. Hence, to complete the proof we need to show that $\barbR_{\alpha}^{-1}+\barbC_{\alpha} \succeq \bbzero$ so that the aforementioned lemma can be invoked.
To that end, suppose first that $\alpha<0$. Then we have that $\barbR_{\alpha} \succ \bbzero$ and $\barbC_{\alpha}=\alpha^{-1} \diag(\bbc)\preceq \bbzero$, so that $\barbR_{\alpha}^{-1}+\barbC_{\alpha} \succeq \bbzero$ may not be positive definite. Suppose now that $\alpha>0$. Then $\barbC_{\alpha} \succeq \bbzero$ and there always exists a sufficiently small positive $\alpha$ such that $\barbR_{\alpha} \succ \bbzero$ since $\bbR_{w} \succ \bbzero$. This implies that if  $\alpha$ is chosen such that $\alpha > 0$ and $\barbR_{\alpha} =\bbR_{x}+\bbR_{w} - \alpha \bbI_{n} \succ \bbzero$ (i.e., the conditions stated in the proposition), then $\barbR_{\alpha}^{-1}+\barbC_{\alpha} $ is positive definite and problems \eqref{eqn_prob1b_forward} and \eqref{eqn_prob2_forward} are equivalent.\hfill$\Box$

\subsection{Proof of Proposition \ref{prop_hsampling_forward}} \label{proof_prop_hsampling_forward}

%
	Defining matrix $\barbC=\diag(\bbc)$, the estimated output is 
	
	\begin{equation} \nonumber
	\hby = \bbH \bbC^{T} \bbC (\bbx+\bbw) = \bbH \diag ( \bbc ) (\bbx+\bbw)=\bbH \barbC (\bbx+\bbw).
	\end{equation}
	Since the desired output is $\bby=\bbH \bbx$, the MSE simplifies to
	
	\begin{equation*}
	\E { \| \bbH \bbx - \bbH \barbC(\bbx+\bbw) \|_{2}^{2} }  = \tr \left[ \bbH \bbR_{x} \bbH^{T} - 2 \bbH \barbC \bbR_{x} \bbH^{T} + \bbH \barbC (\bbR_{x}+\bbR_{w}) \barbC \bbH^{T} \right]. 
	\end{equation*}
	Introducing the auxiliary variable $\bbY \in \reals^{m \times m}$, $\bbY \succeq \bbzero$, minimizing the MSE with respect to $\bbC \in \ccalC$ is equivalent to solving
	
	\begin{align}
	\min_{\substack{\bbc \in \{0,1\}^{n},\\\bbY, \barbC}}\
	& \tr \left[ \bbY \right] \label{eqn_prob1b_hsampling_forward} \\
	\st\
	&\barbC = \diag(\bbc) \ , \ \bbc^{T} \bbone_{n}=p \nonumber \\
	&  \bbH \bbR_{x} \bbH^{T} - 2 \bbH \barbC \bbR_{x} \bbH^{T} 
	+ \bbH \barbC (\bbR_{x}+\bbR_{w}) \barbC \bbH^{T} \preceq \bbY. \nonumber
	\end{align}
	Finally, using the Schur complement-based lemma for positive semidefiniteness~\cite{bova04}, then  \eqref{eqn_prob1b_hsampling_forward} can be shown equivalent to \eqref{eqn_prob2_hsampling_forward}, competing the proof.\hfill$\Box$
%

%
\section*{References}

\bibliographystyle{IEEEtran}
\bibliography{myIEEEabrv,bib-sketching}

\end{document}